\documentclass[aps,pre,twocolumn,amsmath,amssymb,nofootinbib,floatfix,superscriptaddress]{revtex4-2}

\usepackage{amsmath}
\usepackage{amssymb}
\usepackage{amsthm}
\usepackage{amsfonts}
\usepackage{listings}
\usepackage{physics}
\usepackage{enumerate}
\usepackage{latexsym}
\usepackage{psfrag}
\usepackage{bm}
\usepackage{graphicx}
\usepackage[caption=false]{subfig}
\usepackage{blkarray}
\usepackage{array}
\usepackage{color}
\usepackage[normalem]{ulem}
\usepackage{hyperref}
\usepackage[table]{xcolor}
\usepackage{colortbl}
\usepackage{tabularx,diagbox}
\hypersetup{
     colorlinks = true,
     linkcolor = magenta,
     citecolor = magenta
}
\usepackage{mathtools}

\newcolumntype{Y}{>{\centering\arraybackslash}X}

\allowdisplaybreaks

\begin{document}
\setlength{\tabcolsep}{0pt}
\title{A unified model for linear responses of physical networks
} 

  \author{José M. Ortiz-Tavárez}
   \affiliation{
 Department of Physics,
  University of Michigan, Ann Arbor, 
 MI 48109-1040, USA
 }
  \author{William Stephenson}
   \affiliation{
 Department of Physics,
  University of Michigan, Ann Arbor, 
 MI 48109-1040, USA
 }
  \author{Xiaoming Mao}
 \affiliation{
 Department of Physics,
  University of Michigan, Ann Arbor, 
 MI 48109-1040, USA
 }
  \affiliation{Center for Complex Particle Systems (COMPASS), University of Michigan, Ann Arbor, USA}

\begin{abstract} 
Many physical systems---from mechanical lattices and electrical circuits to biological tissues and architected metamaterials---can be understood as networks transmitting \emph{physical quantities}. We present a unified mathematical framework for describing linear responses of such physical networks using tools from algebraic graph theory. This approach captures static and dynamic behaviors across multiple domains, including mechanical, electrical, thermal, and diffusive responses using node and edge variables (e.g., potentials, flows). Our formalism connects multiscale and multi-domain responses to the underlying network structure. We demonstrate how this framework enables efficient, generalizable solutions to a wide class of linear response problems, including stress propagation, charge transport, and wave dynamics, and provide insights into network duality and entropy production.
\end{abstract}
\maketitle

\section{Introduction}
All materials are fundamentally discrete at the atomic scale.  As one zooms out to larger scales, microscopic heterogeneities often average out, and a homogenized continuum theory emerges as a good description of physical properties~\cite{cioranescu1999introduction,torquato2002random,milton2003theory}. For instance, in crystals, long-range order based on repeating unit cells enables elegant mathematical formulations, such as band theory. In glasses, despite the lack of long-range order, statistical rotational and translational invariance beyond a few atomic lengths allows for the construction of effective continuum theories grounded in symmetry principles.

However, many materials exhibit structural complexity across multiple length scales and cannot be captured by traditional homogenization alone~\cite{lakes1993materials,park2007biomaterials,mao2024complexity}. A prime example is biological tissue, which incorporates hierarchical organization from macromolecules and organelles to cells and vascular networks, spanning nanometers to millimeters~\cite{park2007biomaterials}. Such multiscale materials are also ubiquitous in engineering---fiber-reinforced composites, architected metamaterials, and hierarchical porous media---where structural features span from nanometers to centimeters to achieve tailored mechanical, thermal, or acoustic properties~\cite{lakes1993materials,park2007biomaterials,mao2024complexity}.
Modeling such multiscale architectures remains a grand challenge, as conventional multiscale modeling methods are often tailored to specific materials and difficult to generalize across different material classes~\cite{lukkassen2000hierarchical,bouvard2009review,weinan2011principles,ramirez2019homogenized}.  

In parallel, advances in network science offer a powerful alternative perspective~\cite{barabasi2013network,newman2018networks}. Can we characterize the architecture of multiscale materials statistically---using concepts such as connectivity, community structure, geodesic distance, centrality, and spectral properties---to build efficient and generalizable theoretical models? Such an approach holds promise for capturing the essential features of complex materials where structural details vary widely but exhibit statistical regularities. Recent developments in this direction have yielded exciting new insights into systems ranging from granular matter and gels to geological materials~\cite{bassett2012influence,zhang2017fiber,papadopoulos2018network,kollmer2019betweenness,dehmamy2018structural,zhang2019correlated,rocklin2021elasticity,machlus2021correlated,vecchio2021structural,yang2024graph,pete2024physical}.

In this paper, we provide a mathematical framework to describe linear responses of physical networks, which can serve as a unifying tool to accelerate discoveries in this  new front.  Discrete network-based models arise in many contexts in physics and engineering, from Maxwell's mechanical frames and phonons in atomic lattices~\cite{lubensky2015phonons,mao2018maxwell}, to electric circuits~\cite{Blackwell,Bollobás1998} and finite element modeling of multiphysics models~\cite{szabo2021finite,zimmerman2006multiphysics}, offering a rich foundation of knowledge.  Building on this, we extract universal elements from these approaches and propose a streamlined mathematical framework grounded in algebraic graph theory~\cite{swamy1981graphs,biggs1993algebraic}, capable of representing diverse physical phenomena, static and dynamic, in a unified language. 
As summarized in Table~\ref{mapping}, our approach maps linear responses in various domains---mechanical, electrical, thermal, diffusive---to \emph{potentials and flows on nodes and edges} of a network.  This mapping, combined with the linear algebraic structure of graphs that relates nodes and edges, enables an efficient and extensible framework for analyzing physical responses, coupling across domains, and connecting physical behavior to underlying network architecture.

\begin{table*}[t!]
\centering
\setlength{\extrarowheight}{0pt}
\renewcommand{\arraystretch}{1.5}
\begin{tabularx}{\textwidth}{|>{\centering\arraybackslash}p{3cm}|Y|Y|Y|Y|Y|}
\hline
\cellcolor{gray!50}\textbf{Properties}\phantom{\Bigg|}  & \cellcolor{blue!25}\textbf{Mechanical (static)} & 
\cellcolor{blue!25}\textbf{Mechanical (mobility)} &\cellcolor{yellow!30}\textbf{Electrical} & \cellcolor{red!30}\textbf{Thermal} & \cellcolor{green!30}\textbf{Diffusive} \\
\hline
\cellcolor{gray!20} Potential  & Position $\mathbf{U}_i$ & Velocity $\mathbf{V}_i$ & Electric potential $V_j$ & Temperature & Chemical potential  \\
\hline
\cellcolor{gray!20}Node incoming current & Force $\mathbf{F}_i$ & Force $\mathbf{F}_i$ & Incoming current $I_j$ & Incoming heat & Incoming matter \\
\hline
\cellcolor{gray!20} Potential difference  & Extension $e_{nm}$ & Extension rate $\dot{e}_{nm}$ & Voltage $v_{nm}$ & Temperature difference & Chemical potential difference \\
\hline
\cellcolor{gray!20}Flow   & Tension $t_{nm}$ & Tension $t_{nm}$ & Current $i_{nm}$ & Heat current & Matter current \\
\hline
\cellcolor{gray!20}Transported quantity & Momentum & Momentum & Charge & Heat & Matter \\
\hline
\cellcolor{gray!20} Conductivity & Stiffness  & Mechanical impedance  & Electrical admittance & Thermal conductivity & Diffusivity \\
\hline
\end{tabularx}
\caption{Analogous potentials and flows on physical networks across different domains.}\label{mapping}
\end{table*}

This paper is organized as follows.  In Sec.~\ref{SEC:physical} we introduce the basic formulation of how linear responses in multiple domains (Table~\ref{mapping}) can be written in a unified language based on potential-flow problem on networks. In Sec.~\ref{SEC:static} we discuss static responses of mechanical networks to showcase the utility of this method in efficiently calculating a diverse set of problems.  In Sec.~\ref{SEC:irreversible} we discuss a set of irreversible transport problems on networks, the formulation of which runs in parallel with static problems, albeit with one time derivative. In Sec.~\ref{SEC:waves} we discuss dynamics and wave propagation on networks.  

\section{Physical networks}\label{SEC:physical}

\subsection{Notation}\label{SEC:notation}
First we  define some conventions of notation that we will use in this paper.  Because both network indices and Cartesian indices are necessary for mechanical networks, we adopt a unified convention of notation here. 

Upper indices  identify columns while lower indices identify rows. Repeated indices are summed over only when they appear as both lower and upper indices. 
Lower case Latin indices label nodes. They can also label cycles or faces.
Greek indices label spatial directions $x,y,z$. 
Upper case Latin indices label irreversible processes. These are only used in the Sec.~\ref{SEC:irreversible}.

We  often group indices. For example $e_{ij}$ represents a quantity on edge $ij$. Here $ij$ plays the role of an edge index and is understood to run only over edges that actually exist in the network. We adopt this notation of using the indices on the two nodes an edge connects to label the edge so it naturally denotes a directionality from $i$ to $j$, which is useful for the discussion of physical transport. 

Consider $U_n^\alpha$ in a two dimensional network of three nodes. The lower index $n$ labels a row while the upper index $\alpha$ labels a column. Written explicitly as a matrix we have:
\begin{equation}
    U_n^\alpha \equiv
    \begin{pmatrix}
        U_1^x & U_1^y \\
          U_2^x & U_2^y \\
            U_3^x & U_3^y \\
    \end{pmatrix}.
\end{equation}

If we instead write $U_{n\alpha}$, then $n\alpha$ labels a row:
\begin{equation}
    U_{n\alpha} \equiv
    \begin{pmatrix}
        U_{1x} \\
        U_{1y} \\
          U_{2x} \\
          U_{2y} \\
        U_{3x} \\
          U_{3y} \\
    \end{pmatrix}.
\end{equation}

\subsection{Flows and potentials on networks}\label{SEC:electric}

We use the term physical network to refer to structures where there occurs a transmission or flow of some physical quantity through the edges, and said flow can lead to accumulation on the nodes. Examples of this transmitted quantity or flow include electric current, thermal or diffusive flow, and mechanical tension, among others. 
This is in contrast to non-physical networks, such as social networks, where connections capture relationships and are not necessarily concerned with the flow of any quantity~\cite{newman2018networks}.

In this paper we will describe a large set of physical problems in terms of potentials and flows. This general approach was first developed in the context of control theory where flows are referred to as the ``through'' variables and potential differences as the ``across'' variables \cite{Blackwell}. While this literature mainly focused on electric and pipe flow networks as scalar problems and explored dynamics of these networks, other studies used similar graph theory approaches for static mechanical networks as vector potential problems, exploring zero modes, states of self stress, and duality~\cite{SHAI2001343}.  Here we take our present language from previous work on mechanics~\cite{SHAI2001343}, but aim at providing a comprehensive approach that combines both fields, and discuss a unified frame to solve similar types of linear response problems, static and dynamic. 

\begin{figure}[h!]
        \centering        \includegraphics[width=0.35\textwidth]{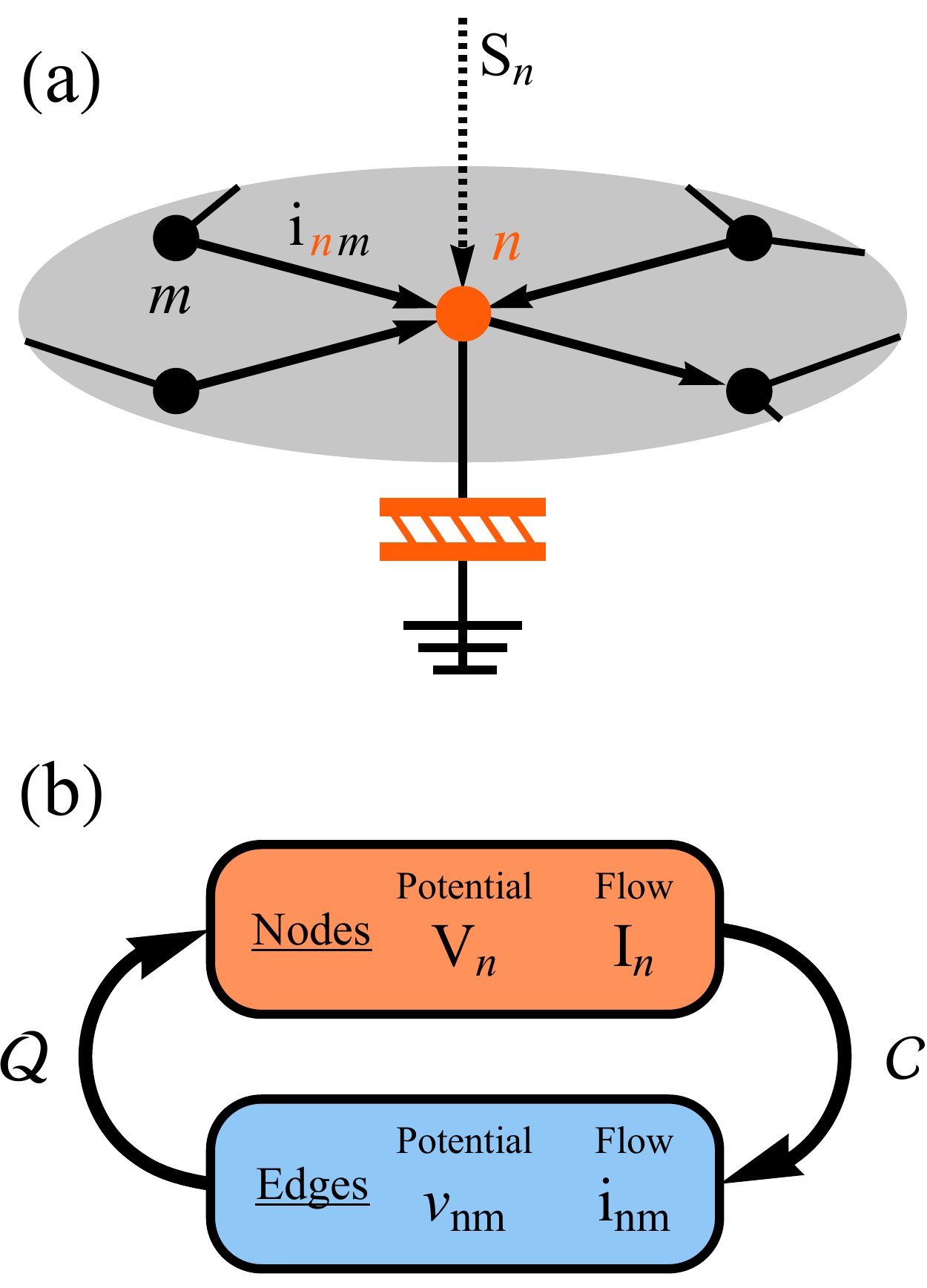}
        \caption{Nodes and edges of a physical network. (a) Net flow to a node, illustrating to Eq.~\eqref{EQ:QI}. (b) Potential and flow feedback between nodes and edges, illustrating the effect of the cut-set ($\mathcal{Q}$) and incidence ($\mathcal{C}$) matrices from Eqs.~\eqref{EQ:electricall}.}
        \label{fig:QC}
\end{figure}

To introduce the basic formulation, let us for concreteness consider the example of transport on an electrical network (Fig.~\ref{fig:QC}a).  The flow of charge through edge $nm$ is the current $i_{nm}$.   Since charge is a conserved quantity the currents must satisfy the discrete continuity equation:
\begin{equation}\label{EQ:QI}    \dot{Q}_j=I_j=\mathcal{Q}_{j}^{nm}i_{nm} + S_j ,
\end{equation}
where $Q_j$ is the amount of charge at node $j$, $I_j$ is the net current flowing into node $j$, $i_{nm}$ is the  current flowing from node $m$ to node $n$, and $S_j$ is an external source term. The ``cut-set'' matrix (see more discussions in Sec.~\ref{SEC:Kirchhoff}): 
\begin{equation}\label{EQ:Q}
\mathcal{Q}_{j}^{nm} \equiv \delta_{j}^{n} -\delta_{j}^{m}  ,
\end{equation}
adds the current on all edges incident on node $j$ (Fig.~\ref{fig:QC}b). (Note that that  $nm$ is a single index for an edge here. If $nm$ is not an edge, it simply does not appear as a row of the matrix.)
This equation is directly analogous to the  continuity equation in continuous space:     
\begin{equation}
        \dot{\rho}=- \nabla \cdot \mathbf{J} +  s,
\end{equation}
where $s$ is a source term, $\mathbf{J}$ is the current density and $\rho$ the charge density.

At the same time, we have electric potentials $V_j$ on the nodes leading to potential differences across edges $v_{nm}$:
\begin{equation}\label{EQ:Cv}
    v_{nm}= \mathcal{C }_{nm}^{j} V_j + v_{0,nm},
\end{equation}
where the matrix $\mathcal{C }_{nm}^{j}$, generally called the incidence matrix of a graph, projects potentials from nodes to edges (i.e., calculates the potential drops on edges as $v_{nm}=V_n-V_m$). $v_{0,nm}$ is an externally imposed potential drop or electromotive force (EMF), for example, from a battery on the edge.  The two matrices are related via
\begin{equation}
    \mathcal{Q}=\mathcal{C}^T.
\end{equation}
(Using our index notation $\mathcal{Q}_{j}^{nm}=\mathcal{C}_{nm}^{j}$.) In this work we  analyze linear physical networks meaning that the flow through and edge is linearly related to a potential difference across the edge via a constitutive relation. Edges can generally be seen as elements, in an electrical context this would be resistances capacitors or inductors. The elements on an edge determine its admittance $y_{nm}$. We then have the constitutive relation on the edges:
\begin{equation}\label{EQ:EdgeElectric}
     i_{nm} =y_{nm} v_{nm}.
\end{equation}
In the case of that element being a resistor, Eq.~(\ref{EQ:EdgeElectric}) is simply Ohm's Law, where the admittance is one over resistance $y_{nm}=-1/r_{nm}$.  Note that flows and potential differences are directed quantities ($v_{nm}=-v_{mn}$), in contrast to the admittance or its analogues which are undirected ($y_{nm}=y_{mn}$). For more on sign conventions see the appendix (App.~\ref{SEC:AppA}).  We can also consider components on the nodes, this is equivalent to having an edge from a node to the ground.  The constitutive relation on the nodes is then essentially identical:
\begin{equation}\label{EQ:NodeElectric}
    I_j = Y_j V_j.
\end{equation}
With Eqs.~(\ref{EQ:QI},\ref{EQ:Cv},\ref{EQ:NodeElectric},\ref{EQ:EdgeElectric}) we have a complete set of equations for the electric network, characterizing the transmission of charges on the network and the responses of the nodes and edges. We can write them together as:
\begin{align}\label{EQ:electricall}
\textrm{edge}\to\textrm{node: }  
&\quad I_j=\mathcal{Q}_{j}^{nm}i_{nm} + S_j \nonumber\\
\textrm{node}\to\textrm{edge: }  &\quad v_{nm}= \mathcal{C }_{nm}^{j} V_j + v_{0,nm} \nonumber\\
\textrm{node: } &\quad I_j = Y_j V_j \nonumber\\
\textrm{edge: } &\quad i_{nm} =y_{nm} v_{nm}.
\end{align}
We will use this electric network problem as our first example to discuss basic algebraic graph theory tools for physical networks.

The same formulation applies to parallel problems in thermal and diffusive transport problems, as listed in Table~\ref{mapping}, which will be discussed in details in the following sections. 

Eqs.~\eqref{EQ:electricall} form a closed set of equations to solve linear response problems in physical networks.  One can write them together in terms of an ``equation of motion on the network'':
\begin{equation}\label{EQ:EOM}
    Y_j V_j= \mathcal{D}_{j}^{i}V_i + \mathcal{Q}_{j}^{nm} y_{nm} v_{0,nm} +S_j ,
\end{equation}
where we have introduced the dynamical matrix 
\begin{equation}
\mathcal{D}_{ji}\equiv \mathcal{Q}_{j}^{nm} y_{nm}  \mathcal{C }_{nm}^{i} .
\end{equation}
We consider $V_j$, potentials on the nodes, as our fundamental degrees of freedom, and this equation solves the responses of the network with respect to external drive on nodes  current sources $S_j$ or on edges voltage sources $v_{0,mn}$.   

In the simple case of $y_{mn}=1$ for all edges, the dynamical matrix reduces to the graph Laplacian:
\begin{equation}
 \mathcal{D}_{j}^{i} \to \mathcal{L}_j^i,
\end{equation}
where the graph spectrum determines the dynamics of the physical problem.  Here we present a more general scenario, where the nodes and edge admittance, $Y_j, y_{nm}$ exhibit their own time scales, and can lead to rich dynamics not captured by the graph Laplacian~\cite{swamy1981graphs,biggs1993algebraic}.  

\subsection{Kirchhoff's laws and compatibility}\label{SEC:Kirchhoff}
In this section we review Kirchhoff's laws~\cite{Kirchhoff_German,Kirchhoff_English} and their relation to basic concepts from graph theory, cycles and cut sets, and the fundamental theorem of linear algebra~\cite{fundamental_theorem_linalg}. This discussion is the basis for the universal applicability of Kirchhoff's laws beyond electrical circuits and also gives a deeper and more expansive understanding of them.

We first start from the condition of current conservation, Kirchhoff's current law, meaning that for all subgraphs the flow coming in equals the flow coming out.  Note that this is assuming there are no sources and no accumulation on the nodes which in the context of  Eq.~\eqref{EQ:QI} means $S_i=0,I_i=0$.
To make this statement mathematically precise we need to define \emph{cut sets}~\cite{swamy1981graphs,biggs1993algebraic}. A cut set is a set of edges that if removed divide the graph into two subgraphs now mutually disconnected.  A flow $i_{nm}$ is conserved if the sum of the flows along every cut set is zero:
   \begin{equation}
       0=\sum_{nm\in \,\textrm{cut set}} i_{nm}  \label{eq:k_currentlaw}.
   \end{equation}
This is Kirchhoff's current law, also called the flow law. Cut sets can be represented as vectors and form a linear subspace called the \emph{cut space}. A basis for this space is called a cut basis. If the flows satisfy the flow law for all elements of a cut basis they satisfy it for any cut set, since any cut set is a linear combination of the basis vectors (shown in  App.~\ref{SEC:AppA}). A matrix such that its rows make a complete cut basis is called a cut-set matrix. A convenient choice for the cut-set matrix is $\mathcal{Q}=\mathcal{C}^T$, since each cut set in the basis is simply the set of edges incident on a node, as we used in Eq.~\ref{EQ:Q}.  In terms of this matrix Kirchhoff's current law reads:
\begin{equation}
    0=\mathcal{Q}_{j}^{nm} i_{nm}.
\end{equation}
Here the label $j$ formally labels a cut set but it also labels a node since for our chosen basis each cut set is identified with one node.  This equation implies that all flows that satisfy the flow law are orthogonal to the cut space. Furthermore, the space of flows that satisfy the flow law is the orthogonal complement of the cut space \cite{swamy1981graphs,Lloyd1978GraphTW}.

\begin{figure}[h!]
        \centering
        \includegraphics[width=0.45\textwidth]{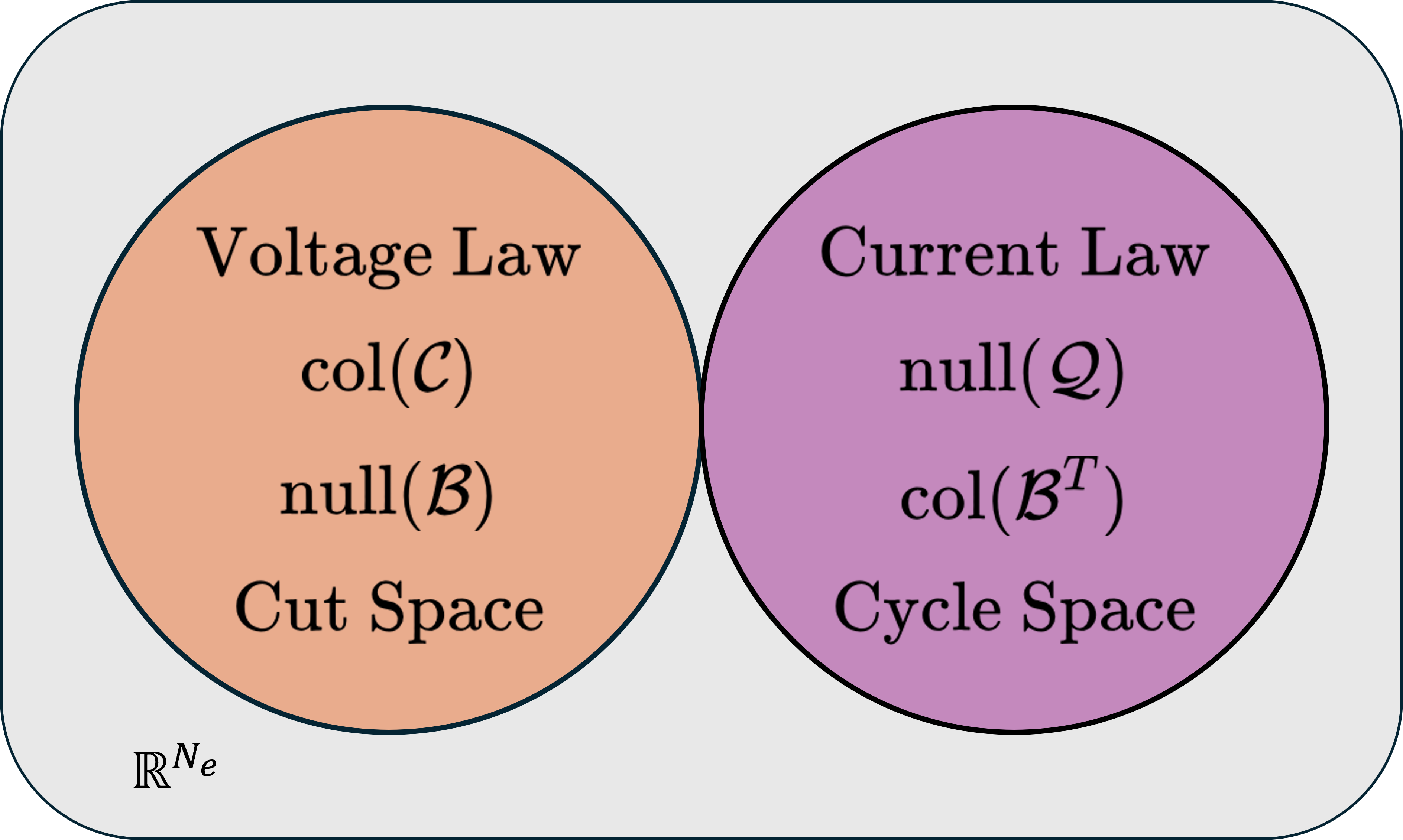}
        \caption{A Venn diagram representation of  vectors on $\mathbb{R}^{Ne}$, where $N_e$ is the number of edges. The space of vectors that satisfies the voltage law is the orthogonal complement of the space of vectors that satisfy the current law and are therefore shown as sharing only one point, the zero vector. All the descriptions within the circles are equivalent. The area outside of both circles but in the box represent vectors that contain components from both subspace.}
        \label{fig:linsubs}
\end{figure}

Given a potential $V_i$ for each node we can compute potential differences $v_{ij}=V_i-V_j$ across each edge. The incidence matrix previously introduced performs this operation:
\begin{equation}
    v_{nm} = \mathcal{C }_{nm}^{i} V_i .
\end{equation}
Note that this matrix has as many rows as there are edges and as many columns as there are nodes.  In general a directed quantity on the edges that can be obtained as potential differences are called \emph{compatible}. It is easy to show that for a given set of  directed quantities $v_{nm}$ there exist a potential such that this flows are obtained as the potential differences if and only if they satisfy the cycle law, which is to say they add up to zero along cycles:
\begin{equation}
    0=\sum_{nm\in \textrm{cycle}} v_{nm} \,\,\forall\, \textrm{cycles of } G .
    \label{eq:circuitlaw}
\end{equation}
This is Kirchhoff's voltage law. Cycles make  a linear subspace which means any linear combination of cycles is also a cycle, in the sense that it satisfies Eq.~\eqref{eq:circuitlaw}. One can always find a set of linearly independent cycles such that all cycles are linear combinations of them, a complete cycle basis. A matrix such that its rows form a cycle basis is called the cycle or circuit matrix $\mathcal{B}$ \cite{SHAI2001343}. In terms of this matrix Kirchhoff's voltage law is written as:
\begin{equation}
    0=\mathcal{B}_{f}^{nm}v_{nm} .
\end{equation}
The space of all vectors which are potential differences is then the null space of $\mathcal{B}$, which is also the orthogonal complement of the cycle space. Since this space is also the column space of the incidence matrix, one can show via the fundamental theorem of linear algebra~\cite{fundamental_theorem_linalg} that the cycle space and the cut space are orthogonal complements. Therefore, these spaces are equivalent to the space of flows that satisfy the flow law and potential differences respectively (Fig.~\ref{fig:linsubs}). 

\subsection{Vector flow problems}\label{SEC:vec}

Potentials and flows need not be one dimensional or scalar quantities. An example of a case where these quantities are vectors is mechanics. For a spring network in $d$ dimensions the potentials can be the $d$ dimensional positions or the displacements and the flows are generally the tensions of the edges. 

Following the notation we defined in Sec.~\ref{SEC:notation}, Latin indices label network elements like nodes, edges and cycles, and Greek indices  label spatial directions. 
For example, potential $V_i^\alpha$ , which should be interpreted as the $\alpha$ component of the potential on node $i$.  Flows on the edges will be denoted as $f_{ij}^\alpha$, which is the $\alpha$ component of flow $f$ to node $i$ through edge $ij$. 

\begin{figure}
    \centering
    \includegraphics[width=0.45\textwidth]{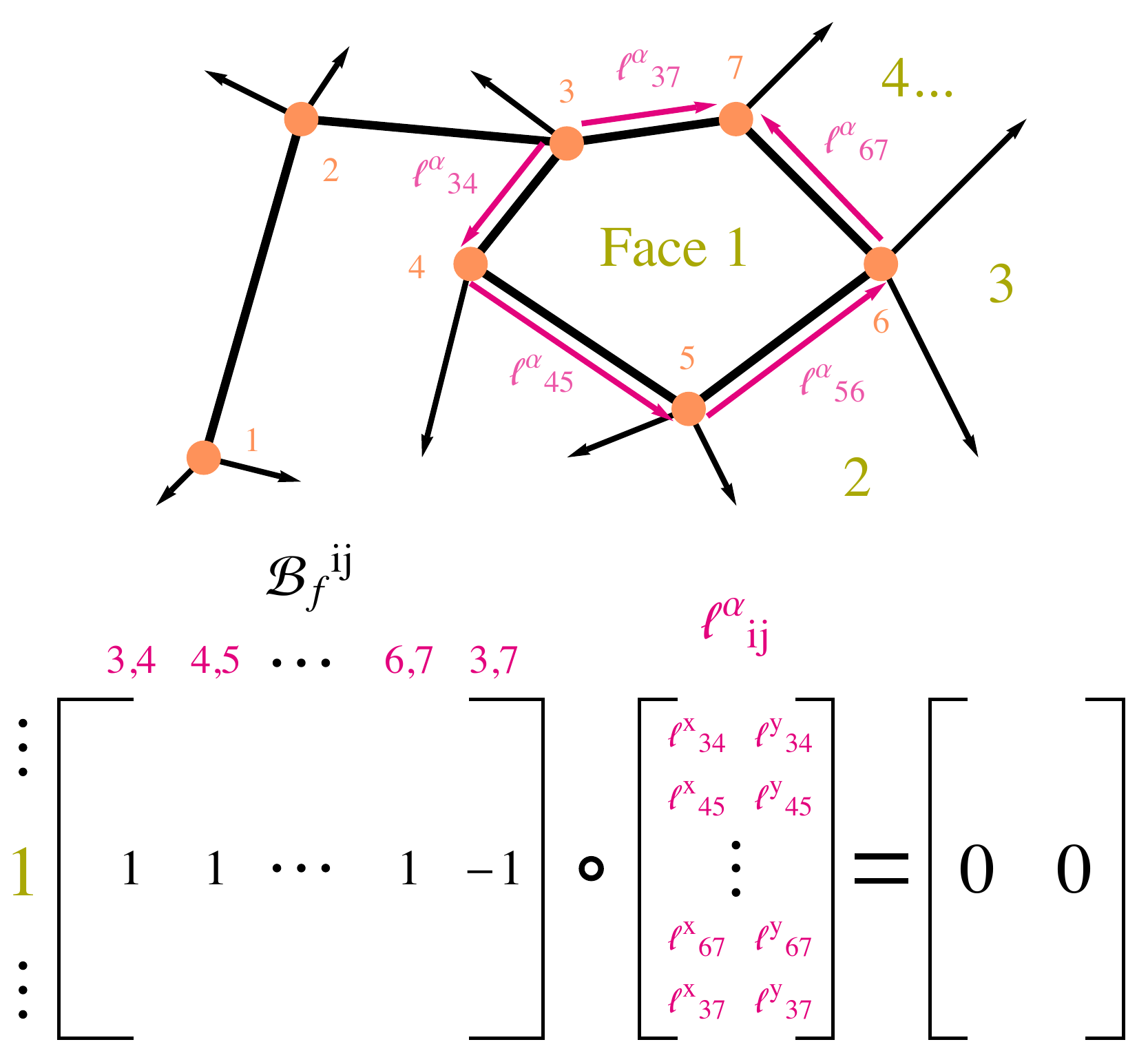}
    \caption{A mechanical network with a closed loop of edges has a non-trivial cycle basis, and therefore a non-trivial cycle matrix ($\mathcal{B}_{f}^{ij}$).
    For a given face (defined by the cycle of edges that surround it), the elements of its row in the cycle matrix will be +1 (-1) in a column if that column corresponds to an edge that points in the direction of (opposite of) a loop around the face, all other elements will be 0.
    The cycle matrix multiplied by the edge length matrix, for a configuration that is embedded in the flat space (i.e., compatible), will result in an all-zero matrix, per Kirchhoff's voltage law (Eq.~\eqref{eq:vector_cycle}).}
    \label{fig:vecvol}
\end{figure}

Kirchhoff's laws discussed previously for scalar quantities apply to the vector quantities component by component. First, the flow law, or Kirchhoff's current law is the force balance condition. Let $t_{mn}^\alpha$ represent the tensions, the force balance condition is then written as:
\begin{equation}\label{EQ:KCM}
    0=\mathcal{Q}_{j}^{mn} t_{mn}^\alpha.
\end{equation}
If we take the node positions as the potentials, then the geometry of the edges described by a vectors  $\ell^\alpha_{ij}=X^\alpha_i-X^\alpha_j$ is a potential difference. The incidence matrix relates potentials to potential differences:
\begin{equation}           
    \ell_{ij}^{\alpha}=\mathcal{C}_{ij}^{k}X_{k}^{\alpha} .
\end{equation}
Kirchhoff's voltage law (potential differences add to zero along cycles) correspond to:
\begin{equation}                    
    0=\mathcal{B}_{f}^{ij}\ell_{ij}^{\alpha}. 
    \label{eq:vector_cycle}
\end{equation}
Here, the index $f$ labels a cycle, as illustrated in Fig.~(\ref{fig:vecvol}) along with a diagrammatic representation of Eq.~(\ref{eq:vector_cycle}). For $\ell^\alpha_{ij}=X^\alpha_i-X^\alpha_j$ from a given geometry, this voltage law is automatically satisfied.  In Sec.~\ref{SubSec:Geometry} we discuss the more general implications of this rule. 

For two-dimensional (2D) planar graphs the set of inner faces makes a complete cycle basis. If one chooses this basis then $f$ labels a face. This is particularly relevant for the study of mechanical duality and graphic statics \cite{SHAI2001343,Zhou2018,Zhou2019}.
 
So far the ``vector problem" of mechanics seems a trivial extension of scalar problems, such as electricity. The crucial difference comes from the constitutive relationship, Hook's law, which is \emph{not} linear. We can write Hooke's law as:
\begin{equation}\label{EQ:nonlineart}
    t_{ij}^\alpha= -k_{ij} \left(  \sqrt{\ell_{ij,\beta} \ell_{ij}^\beta} -  \ell_{0,ij}\right) \,\hat{\ell}_{ij}^\alpha,
\end{equation}
where $k_{ij}$ is the spring stiffness, $\ell_{0,ij}$ is the rest length (distinct from $\ell_{ij}$, the realized length of an edge) and $\hat{\ell}$ is the unit vector that points along the edge, from node $j$ to node $i$. Note that the relationship between potential difference $\ell$, and flow $t$  is not linear in general.  It reduces to linear exactly for one dimensional spring networks and when the rest lengths are zero.  The relationship is effectively linear in the regime of small deformations, aligning the tension to the direction of the edge at the same time. 
In the following subsections we will first present the basic concepts pertaining to this linear regime and the analogy to the scalar version of the problem, taking node displacements $U$ as potentials, and then discuss the general case of geometric compatibility without assuming the linear regime.
  
\subsubsection{Mechanics in the linear regime}
In the linear regime, we are concerned with small, formally infinitesimal, node displacements from some given reference positions $X_k^\alpha \to X_k^\alpha+U_k^\alpha$. Using this in Eq.~\eqref{EQ:nonlineart} and expanding to first order in $U$, we obtain:
\begin{equation}
    t_{ij}^\alpha \simeq -k_{ij} \hat{\ell}_{0,ij}^\beta  (U_{i\beta} -U_{j\beta})
    \hat{\ell}_{0,ij}^\alpha, 
    \label{EQ:tensionlinear}
\end{equation}
where $\hat{\ell}_{0,ij}^\beta  (U_{i\beta} -U_{j\beta})=e_{ij}$ is the length extension of  edge $ij$ (which is $e_{ij}= \ell_{ij}-\ell_{0,ij}$, to first order).
Note that $\hat{\ell}_{0,ij}^\alpha$ is not the orientation of an edge's rest length, but the orientation of the unperturbed edge. 
To first order, $\hat{\ell}_{0,ij}^\alpha$ is equivalent to $\hat{\ell}_{ij}^\alpha$.
For the linear regime, it is then convenient to identify these displacements $U_{i\beta}$, instead of the node positions, as the node potentials, while the flows are still the edge tensions. 
            
The relationship between node displacements and edge elongations is given by the compatibility matrix $C_{ij}^{k\alpha}$: 
\begin{equation}            
    e_{ij}=C_{ij}^{k\alpha} U_{k\alpha},
\end{equation}
which is obtained from the incidence matrix via:
\begin{equation}          
    C_{ij}^{k\alpha}=\mathcal{C}_{ij}^k \hat{\ell}_{0,ij}^\alpha.
\end{equation}
Forces on the nodes are related to tensions via the equilibrium matrix:
\begin{equation}            
    f_{k\alpha}=Q_{k\alpha }^{ij} t_{ij} ,
\end{equation}
which is obtained from the cut-set matrix $\mathcal{Q}=\mathcal{C}^T$ via:
\begin{equation}
    Q_{k\alpha }^{ij}=\mathcal{Q}_{k}^{ij} \hat{\ell}^{ij}_{0,\alpha}.
\end{equation}
Note that we have written the tensions as scalar quantities.  It is clear that $Q=C^T$, a relationship inherited from the cut-set and incidence matrices. This implies then that the column space of $C$ and the null space of $Q$ are orthogonal complements and correspond to compatible elongations and force balanced tensions respectively, a relation used in studies of kinematic and static determinacy of mechanical frames~\cite{CALLADINE1978161,pellegrino1986matrix,calladine1991first,guest2003determinacy}. 

The analogous condition to Kirchhoff's current law (Eq.~\eqref{eq:k_currentlaw})  is force balance. Dropping the indices we can write it as:
\begin{equation}
    0=Q\mathbf{t},
\end{equation}
which is equivalent to Eq.~\eqref{EQ:KCM} using the fact that tensions are along the edges (Eq.~\eqref{EQ:tensionlinear}).
The sets of tensions that satisfy this equation are the states of self stress. See Fig.~(\ref{fig:modes}) for an example. 

The analogous condition to  Kirchhoff's voltage law Eq.~\eqref{eq:circuitlaw} in this linear regime is:
\begin{equation}\label{EQ:Se}
    S \mathbf{e} = 0  \,, \forall \, \mathbf{e}\in \textrm{col}(C),
\end{equation}
where $S$ is a matrix such that its rows form a basis for $\textrm{null}(Q)$, which are the states of self stress~\cite{pellegrino1986matrix}.        
       
Analogously to the scalar case Eq.~\eqref{EQ:electricall} we have the four equations:
\begin{align}
    \textrm{edge}\to\textrm{node: }  
    \quad &F_{\alpha j }=Q_{j\alpha  }^{nm}t_{nm} + F^{\text{ext}}_{j\alpha } \nonumber\\
    \textrm{node}\to\textrm{edge: }  \quad &e_{nm}= C_{nm}^{k\alpha } U_{k\alpha } + e_{0,nm} \nonumber\\
    \textrm{node: } \quad &F_{j\alpha  }= M_j \partial_t^2 U_{j\alpha  } \nonumber\\
    \textrm{edge: } \quad &t_{nm} =-k_{nm} e_{nm}.
    \label{EQ:mechanicall}
\end{align}
Here, forces map to currents and displacements map to voltages. We see that net force on the nodes $F$ takes the role of net current on the nodes $I$. Correspondingly the external force $F^{\text{ext}}$ is the current source term. Identifying force as the flow implicitly identifies momentum as the quantity that is ``flowing," analogously to charge in electric transport. The potential here describes only geometry. The quantity $e_{0,nm}$ corresponds to a small change in the rest length of an edge. The node constitutive relation is simply Newton's second law while the edge constitutive relation is Hooke's law.  This type of formulation has been central in the study of topological mechanics in Maxwell lattices and networks~\cite{sun2012surface,kane2014topological,lubensky2015phonons,sussman2016topological,rocklin2017transformable,mao2018maxwell,xiu2022topological,xiu2023synthetically}.

More generally, the last two equations can include more diverse relations, such as
\begin{align}
    \textrm{node: } \quad &F_{j\alpha  }= M_j \partial_t^2 U_{j\alpha  } + \gamma_j \partial_t U_{j\alpha  } + \mathcal{K}_j U_{j\alpha  },
    \nonumber\\
    \textrm{edge: } \quad &t_{nm} =-k_{nm} e_{nm} -\eta_{nm} \partial_t e_{nm} ,
\end{align}
where $\gamma_j, \mathcal{K}_j$ are the friction coefficient and the pinning spring constant of the node relative to the substrate/matrix, and $\eta_{nm}$ represent the ``loss''  of the edge as a viscoelastic spring. 
        
The dynamical matrix in index notation is: 
\begin{equation}
    D_{p\alpha}^{k\beta }=Q_{p\alpha}^{ij} k_{ij} C_{ij}^{ k\beta} .
\end{equation}
To write this without the indices we define diagonal matrix $K_{ij}^{nm}=k_{ij}\delta_{ij}^{nm}$. Then we rewrite the previous expression as:        
\begin{equation}
    D=Q K C . \label{eq:dmatDef}
\end{equation}

\subsubsection{Geometry, isometries and floppy modes }\label{SubSec:Geometry}

\begin{figure}
    \centering
    \includegraphics[width=0.45\textwidth]{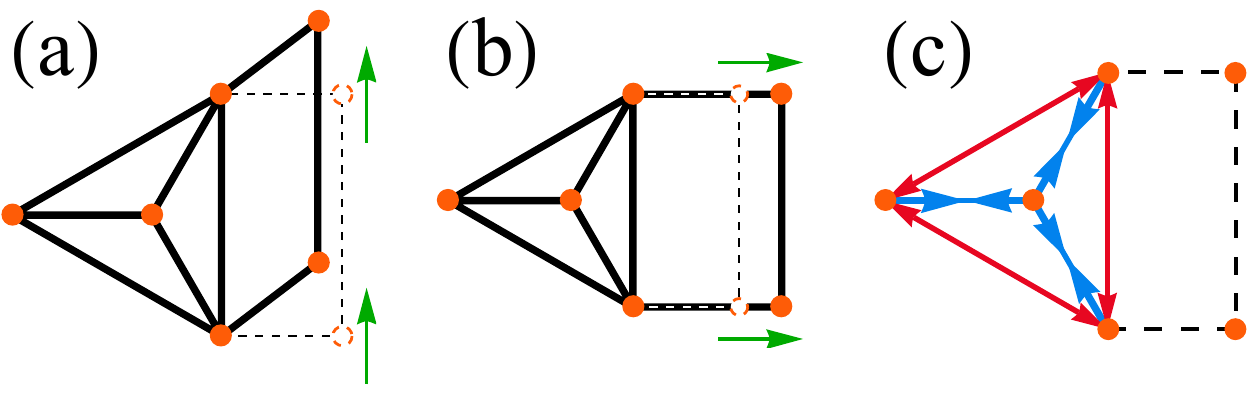}
    \caption{Examples of a floppy mode, where no edge is stretched (a), an isogonal mode, where no hinge rotation occurs at any node (b), and a state of self stress, where edges carry tensions that are force-balanced on the nodes (c).}
    \label{fig:modes}
\end{figure}

We have previously remarked that in the context of mechanics, that the potential describes only geometry.  We  now expand upon this, in a more general way that does not use the linear expansion (Eq.~\eqref{EQ:tensionlinear})

A network embedded in space can generally be described by specifying the graph $G$ (the connectivity), and the position of the nodes $X$, which together form $(G,X)$, a realization of $G$. A realization is simply a way to draw the graph in a given space with all edges as straight lines. This description is common in the field of combinatoric rigidity \cite{asimow1978rigidity}. It is clear that the geometry of the edges $\ell_{ij}^\alpha$  specifies a realization up to a global translation (note that $\ell_{ij}^\alpha$ means all $x,y$ components of each edge are specified).  Therefore the space of edge geometries $\ell_{ij}^\alpha$ that satisfy  Eq.~\eqref{eq:vector_cycle} is equivalent to the space of realizations of $G$.  In other words, each element of this space corresponds to a realization and vice versa,  up to global translation.  This is a vector space which implies that we can add and subtract realizations to each other, and that the result would still be a valid realization. We see here a very basic example of how connectivity constrains geometry. 

Now, consider we are given a graph and the direction of the edges $(G,\hat{\ell})$, but not the length.  
We will now introduce a new matrix, which we shall name the ``closure matrix,'' $B$:
\begin{equation}
    B_{c\alpha}^{ij}= \mathcal{B}_{c}^{ij} \hat{\ell}^{ij}_{\alpha} .
\end{equation}        
The null space of the closure matrix is then: 
\begin{equation}
    B_{c\alpha}^{ij} \ell_{ij}=0 ,
\end{equation}
(for all $c$ and $\alpha$) which gives all possible sets of lengths $\ell_{ij}$ that along with $(G,\hat{\ell})$ yield a valid realization. In other words, all realizations of $G$ such that its edges are parallel to the directions $\hat{\ell}^{ij}_{\alpha}$. The difference between any two such realizations is also an element of this space. These deformations are called isogonal modes, they change the length of the edges but not their orientation in space such that they are angle preserving (Fig.~\ref{fig:modes}, b).  These modes have been discussed in the context of the mechanics of epithelial tissues~\cite{Noll2017}.
         
Although the isogonal modes defined above do not involve linear approximation, they are related to another type of special deformations---zero modes---which are defined in the linear regime.
Zero modes are node displacements that do not elongate any edges (Fig.~\ref{fig:modes}, a). The space of zero modes is the null space of the compatibility matrix. In other words all zero modes correspond to sets of node displacements that satisfy:
\begin{equation}
    0=C\mathbf{U}.
\end{equation}

The term floppy mode is used to denote ``non-trivial" zero modes, the trivial ones being rigid translations and rotations. Floppy modes, along with global rotations can be characterized in terms of the rotation of the edges. 
For two dimensional networks we can obtain the perpendicular component of the elongation as:
\begin{equation}
    e_{ij}^\perp= \hat{\ell}^\beta_{ij} \epsilon_{\beta\alpha}\left(  U_j^\alpha - U_i^\alpha \right) ,
\end{equation}           
where the antisymmetric tensor $\epsilon$ performs a $90^\circ$ degree rotation. We can write this in matrix terms as: 
\begin{equation}
    \mathbf{e}^\perp=C^\perp \mathbf{U} .
\end{equation}
When described in this manner, floppy modes and rotations can be obtained from the closure matrix. The perpendicular component of the elongation corresponds to a zero mode if and only if    
\begin{equation}
    0=B  \mathbf{e}^\perp .  
    \label{eq:floppyclosure}
\end{equation}
            
Therefore, in two dimensions, there is a one-to-one mapping between isogonal modes and zero modes (excluding translations).  Note that the row space of $B$ ($\textrm{row}(B)$) in general is \emph{not} the same as $\textrm{null}(Q)$. The two spaces are equivalent only in the one dimensional case.  Instead, the $\textrm{null}(Q)$ is equivalent to the row space of $\textrm{row}(S)$ defined in the linear regime (Eq.~\eqref{EQ:Se}). 

\subsubsection{Dual graphs and reciprocal diagrams}
It is well established in graph theory that cycles and cut sets of a planar graph $G$ map to cut sets and cycles of its dual graph $\tilde{G}$, which is constructed by mapping faces, nodes, and edges of $G$ to nodes, faces, and edges of $\tilde{G}$~\cite{swamy1981graphs,biggs1993algebraic}. Using the formulation we defined in Sec.~\ref{SEC:physical}, and take cycles in the $\mathcal{B}$ matrix to be faces of the graph, 
this can be written as
\begin{align}\label{EQ:QB}
    \mathcal{Q}(G)=\mathcal{B}(\tilde{G}),
\end{align}
which leads to:
\begin{align}
    \textrm{Flows that satisfy current law of } G &= \textrm{null}(\mathcal{Q}(G))
    \nonumber\\
    &\updownarrow \nonumber\\
    \textrm{Flows that satisfy  voltage law of } \tilde{G}&= \textrm{null}(\mathcal{B}(\tilde{G})).
\end{align}

\begin{figure}[h!]
    \centering        
    \includegraphics[width=0.45\textwidth]{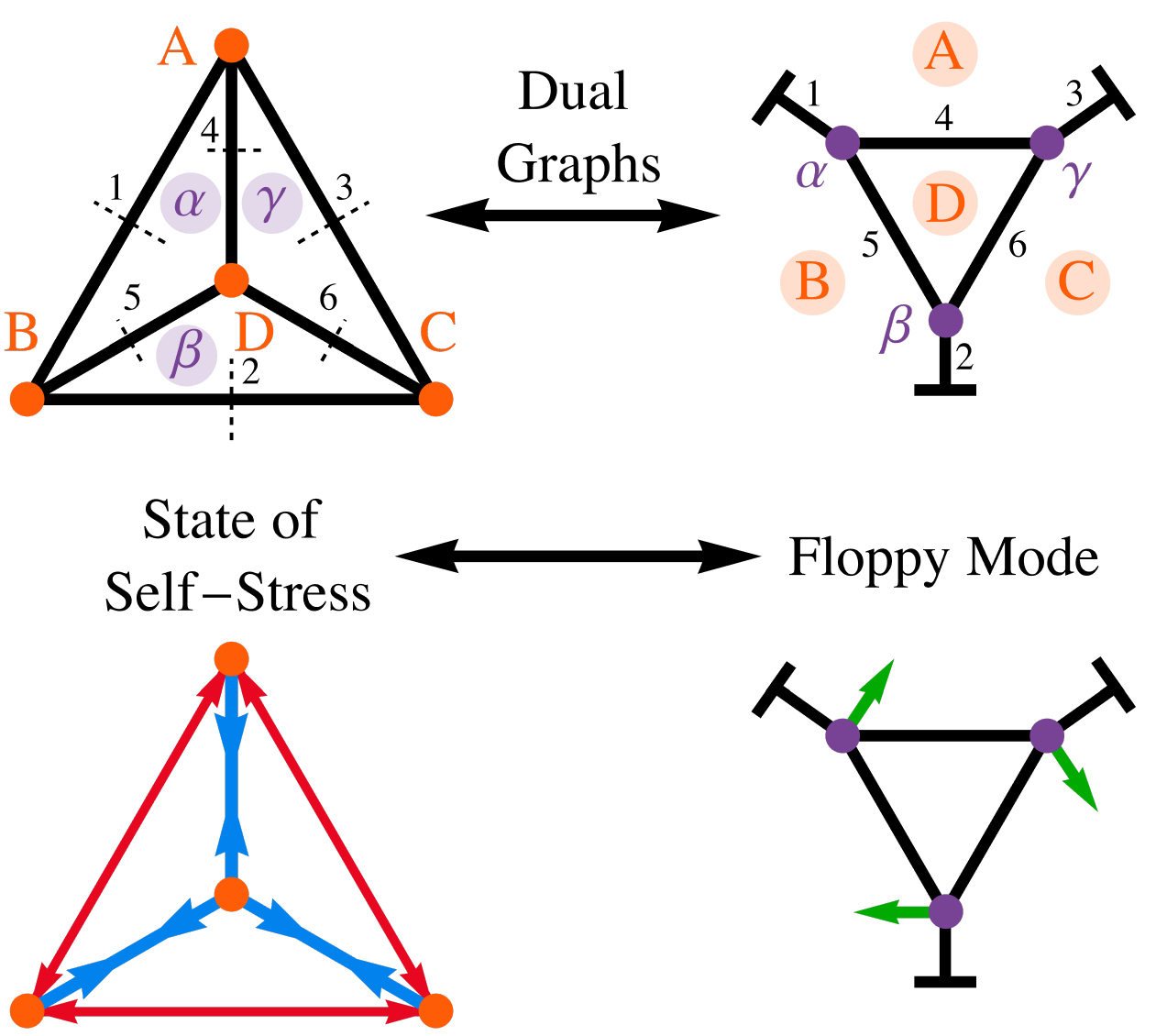}
    \caption{A graph with  $A$-$D$, faces $\alpha$-$\gamma$, and edges $1$-$6$ has a dual graph with faces $A$-$D$, nodes $\alpha$-$\gamma$, and edges still 1-6, but now lying perpendicular to their original direction.
    Exterior edges in the original graph become attached to a pin in the dual.
    The states of self-stress in a graph are mapped to floppy modes (as in this example) or overall rotations in the graph's dual. 
    }
    \label{fig:dualGraph}
\end{figure}

A straightforward extension of this duality relation in the vector problem is the construction of Maxwell-Cremona reciprocal diagrams~\cite{maxwell1864xlv,Maxwell_1870,cremona1890,CALLADINE1978161,crapo1994spaces,mitchell2016mechanisms,mitchell2015mechanisms}, where the dual graph is embedded in the same space, $(\tilde{G},\tilde{X})$ with each edge in the reciprocal diagram parallel to its corresponding edge in the original.  Note that it is more common to rotate the reciprocal diagram by $90^\circ$ such as the corresponding edges are perpendicular (as shown in the example Fig.~\ref{fig:dualGraph}), but it is more convenient to keep them parallel for this discussion. 

The extension of Eq.~\eqref{EQ:QB} is then:
\begin{align}
    Q(G,X)=B(\tilde{G},\tilde{X}),
\end{align}
giving
\begin{align}
    \textrm{States of self stress of } G &= \textrm{null}(Q(G))
    \nonumber\\
    &\updownarrow \nonumber\\
    \textrm{Isogonal modes of } \tilde{G}&= \textrm{null}(B(\tilde{G})),
\end{align}
which further equals non-translational zero modes of $G$ in the linear regime. This duality shows up in various contexts such as force network ensembles of granular matter~\cite{snoeijer2004force,tighe2010force,henkes2009statistical} and topological floppy modes in disordered networks~\cite{Zhou2019}. 

\section{Static responses of mechanical networks}\label{SEC:static}
In this section we  study the linear response of mechanical networks to static forces and displacements. This implies that all of our discussion here is only valid in the ``linear regime" which means close to mechanical equilibrium (Eq.\eqref{EQ:tensionlinear}).  We first discuss the case where this mechanical equilibrium is stress-free (all edges are at their rest lengths), and then generalize to the stressed equilibrium case in Sec.~\ref{SEC:prestress}.  This set of tools can be useful in computational studies of mechanical networks such as jamming and rigidity percolation~\cite{jacobs1995generic,o2003jamming,ellenbroek2006critical,ellenbroek2015rigidity,goodrich2012finite,liarte2019jamming,lerner2016statistics} and metamaterials~\cite{bertoldi2017flexible,ortiz2023assur,PhysRevLett.113.245502}.

The methods described here are also applicable to physical networks which are not mechanical. In a later section we will discuss how this applies to networks of irreversible flows near thermodynamic equilibrium. 
        
Mechanical linear response problems are in general elastic energy minimization problems where there is some imposed quantity that acts as a constraint, displacement, force or edge swelling to give some examples. We start with a spring mass network in an unstressed configuration. For convenience we will define the following matrix:
\begin{equation}
    C'=\sqrt{K}C ,
\end{equation}
where $\sqrt{K}$ is a diagonal matrix where each element of the diagonal is the square root of the corresponding spring stiffness (assumed to be all positive). We can define $Q'=C'^T$. These matrices are analogous to the compatibility  and equilibrium matrices ($C,Q$), just with the spring constants incorporated. We will describe the state of the edges, both elongations and tensions, with a vector $\mathbf{w} = \sqrt{K} \mathbf{e}$ that is related to displacement and forces via:
\begin{equation}
    \mathbf{w}=C'\mathbf{U},
\end{equation}
\begin{equation}
    \mathbf{F}=Q'\mathbf{w} .
\end{equation}
It should be clear that the total elastic energy is $E=|\mathbf{w}|^2/2$.  For this reason we will solve all the following linear response problems by writing an expression for and then minimizing $|\mathbf{w}|$  given the constraints.  We will do this by exploiting the fact that the space of compatible elongations $\textrm{col}(C')$ and the space of self stresses $\textrm{null}(Q')$ are orthogonal complements.
In what follows we will often use the projection operator $S'S'^T$ where  $S'$ is a matrix such that its columns make an orthonormal basis for $\textrm{null}(Q')$ (similar  to Eq.~\eqref{EQ:Se} but for $Q'$).  At other times we will use $SS^T$ where $S$ is a matrix so that its columns make an orthonormal basis for $\text{null}(Q)$. 

It is worth noting that one could always solve the problems via the equations of motion without using this minimization approach. Nevertheless we find that this minimization approach is offers an elegant perspective that deepens and unifies the understanding of the subject.
        
\subsection{Response to forces on nodes}
When a force $\mathbf{F}$ (a vector in the $\mathbb{R}^{Nd}$ space of node degrees of freedom, where $N,d$ are the number of nodes and spatial dimension) is exerted on the nodes, the network responds with node displacements which give rise to edge tensions that balance the imposed force. We have then the following problem. We must minimize $|\mathbf{w}|$ given the constraint that it  balances the external force on the nodes,    
\begin{equation}
      \underset{\mathbf{w}}{\text{min}} \; |\mathbf{w}|^2 \quad \text{subject to } \mathbf{F}=Q'\mathbf{w} . \label{eq:minimization1}
\end{equation}  

\begin{figure}[h]
    \centering
    \includegraphics[width=0.45\textwidth]{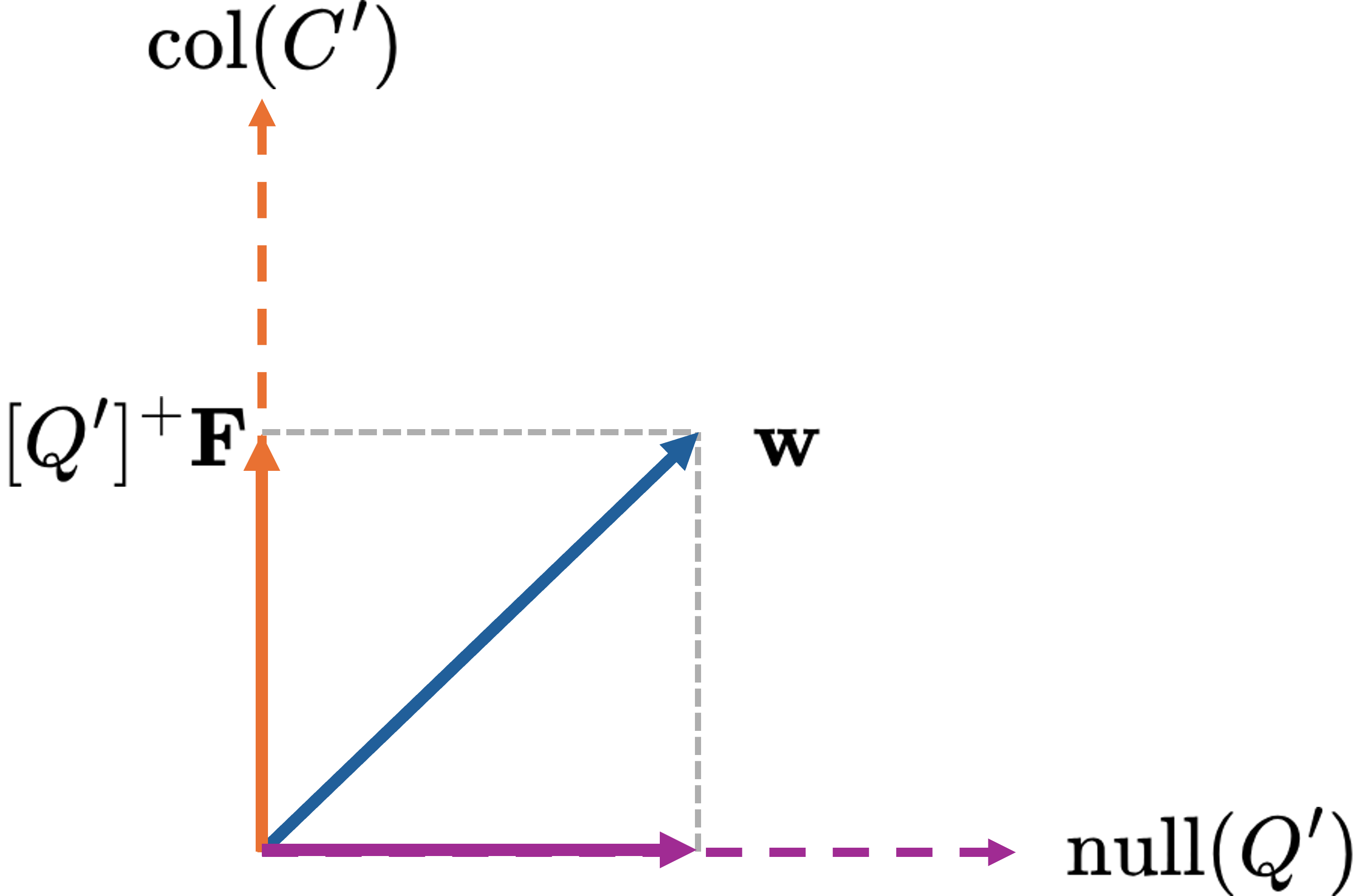}
    \caption{
    The minimization problem in Eq.~(\ref{eq:minimization1}) over $\mathbf{w}$ requires $\mathbf{F}=Q'\mathbf{w}$ in order to balance the applied external force $\mathbf{F}$.
    Therefore, any possible tension response for a given external force must equal $[Q']^+\mathbf{F}$ when projected onto $\text{col}(C')$.
    In this schematic, the valid domain over which we minimize $|\mathbf{w}|^2$ is therefore any point collinear to the horizontal dashed gray line.
    Shown in blue is an arbitrary hypothetical $|\mathbf{w}|$ that meets this condition.}
    \label{fig:force_projection}
\end{figure}

\begin{figure}
    \centering
    \includegraphics[width=0.9\linewidth]{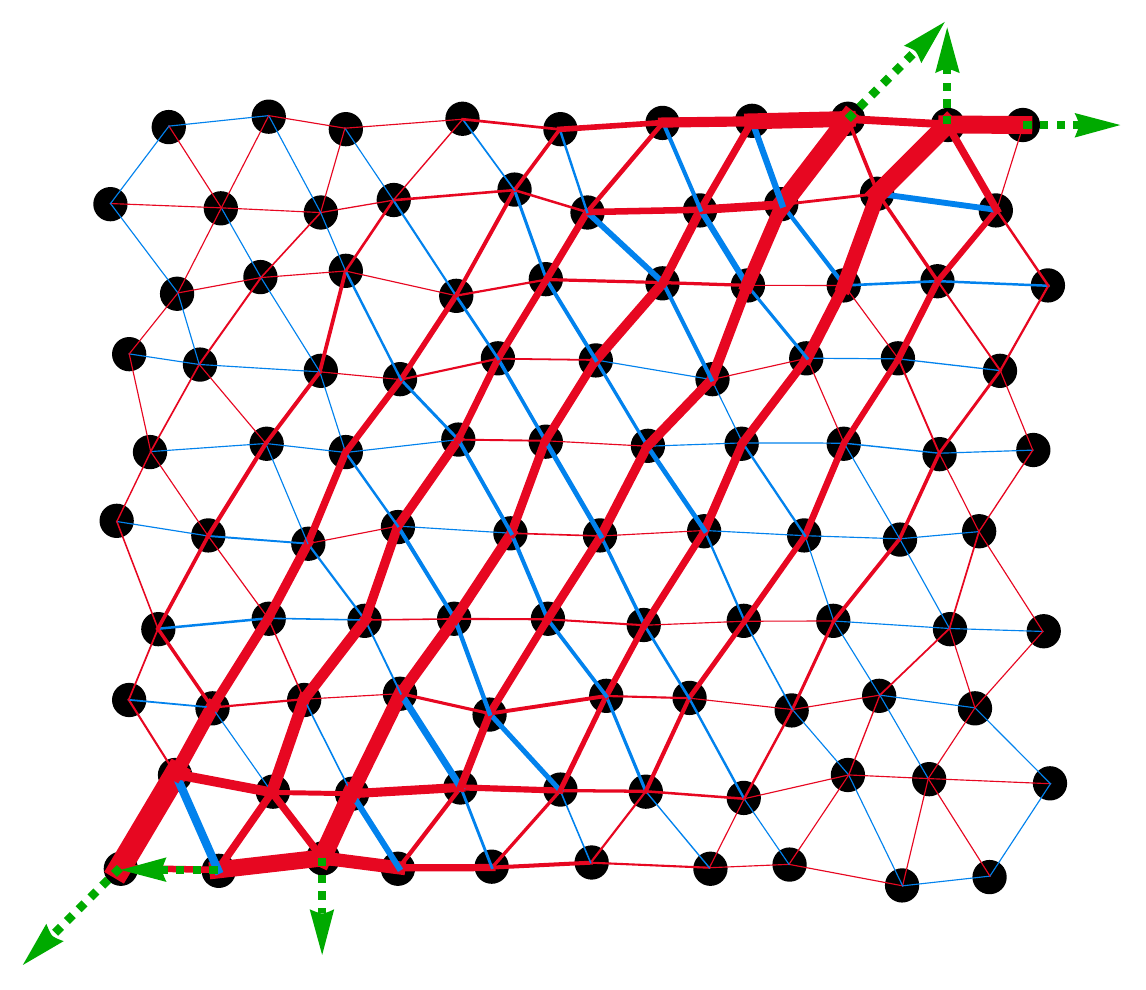}
    \caption{Static forces (green dashed arrows) are applied to a few nodes in the corners of a disordered triangular lattice generated by randomly displacing nodes on a regular triangular lattice by small amounts. 
    The resulting response of the system sees many edges extended (red) and some edges compressed (blue) (thickness proportional to magnitude of tension) in order to for every node to be force-balanced.}
    \label{fig:responsetoforce}
\end{figure}

Any $\mathbf{w}$ can be written as a sum of two components---its projection along the compatible space $\text{col}(C')$ and the self stress space $\text{null}(Q')$ (Fig.~\ref{fig:force_projection}).   
Here we assume that before applying the force the network is unstressed. The network can only realize tensions via node displacements, $ \mathbf{w}_{\text{min}}=C'\mathbf{U}$ for some $\mathbf{U}$.  As a result, $\mathbf{w}_{\text{min}}\in \textrm{col}(C')$.  To see that this $\mathbf{w}_{\text{min}}$ is unique note that the difference between any two $\mathbf{w}$ that result in the same force on nodes has to be an element of $\textrm{null}(Q')$.   This would imply that for any $\mathbf{w}$ that balances the external forces $\mathbf{F}$ its projection on to $\textrm{col}(C')$ is $\mathbf{w}_{\text{min}}$. 
To obtain this projection one can use the Moore-Penrose pseudo inverse~\cite{Penrose_1956}, which we here denote by $+$:
\begin{equation}
    \left[Q'\right]^+ \mathbf{F}=\mathbf{w}_{\text{min}},
\end{equation}
as in Fig.~(\ref{fig:force_projection}).   
We can also solve for node displacements such that $\mathbf{w}_{\text{min}}=C'\mathbf{U}$ by solving:
\begin{equation}
    \mathbf{F}=D\mathbf{U} ,
\end{equation}
where $D$ is the dynamical matrix $D=Q'C'=QKC$. An example of this computation is shown in Fig.~(\ref{fig:responsetoforce}). 
This can also be done by using the pseudo inverse of $D$ or by the use of other numerical methods.

\subsection{Response to edge swelling}
There are circumstances where one can impose a change in the state of the edges by means other than node displacement. One straight forward example is a change in rest-length the springs via some active element, which we call ``swelling'' for simplicity~\cite{rocklin2017directional}, but the actual cause can be diverse, such as piezoelectric coupling, chemical reactions, growth, etc.  

In all of these cases we are producing some change in the edges by accessing some degrees of freedom beyond node displacements, which are normally available for a mechanical network. 
We can describe the imposed change as some $\mathbf{w}_0$, the system will then try to release the resulting stress via node displacements to minimize the energy. The final state of the system is described by
\begin{equation}
    \underset{\mathbf{U}}{\text{min}} \;  |\mathbf{w}|^2 \quad \text{subject to } \;  \mathbf{w}=\mathbf{w}_0-C'\mathbf{U} .\;
    \label{edge_swelling_problem}
\end{equation}                
This is a well known minimization problem in the context of the method of least squares~\cite{fundamental_theorem_linalg}.   
               
\begin{figure}[h]
    \centering
    \includegraphics[width=0.45\textwidth]{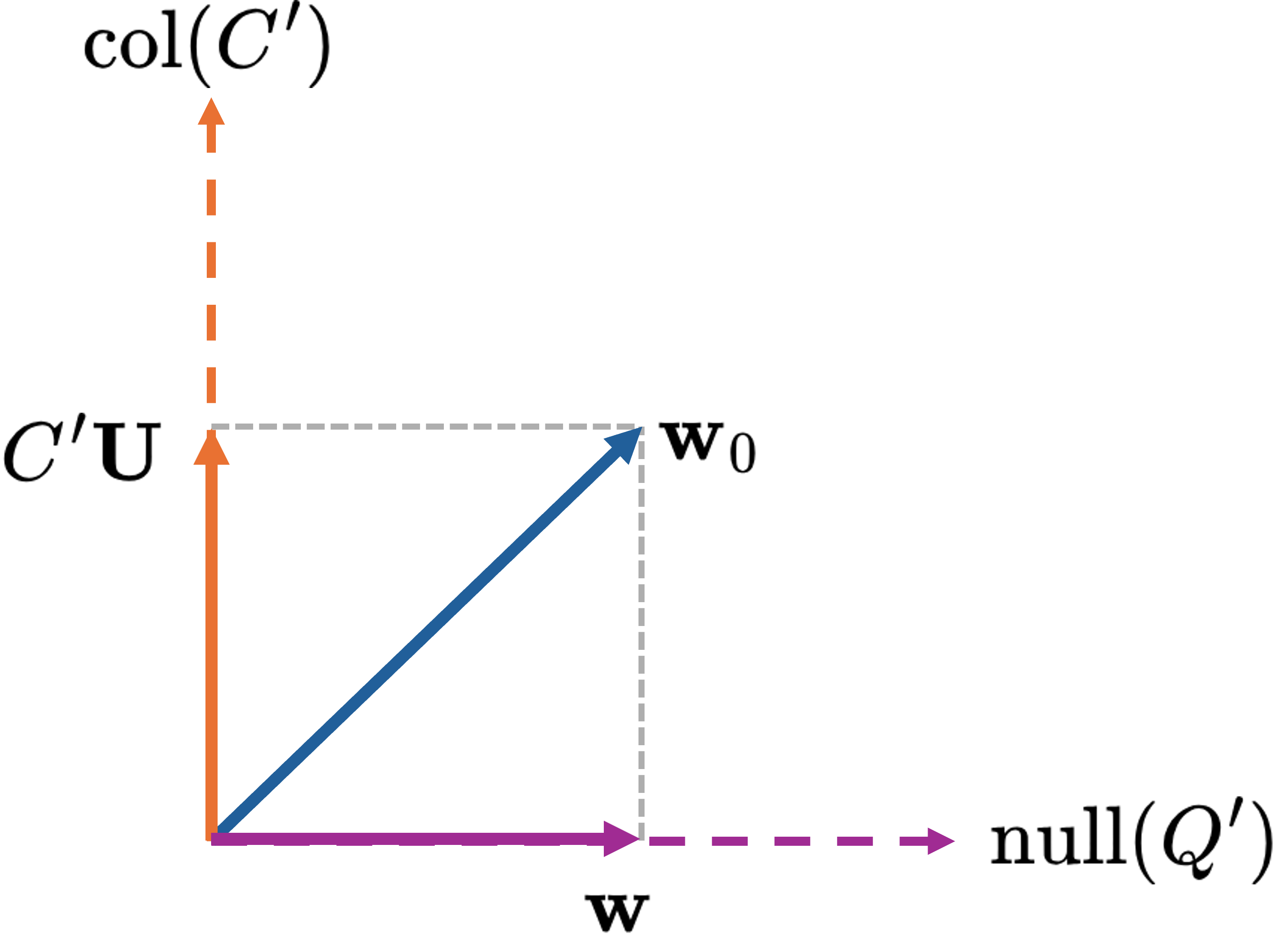}
    \caption{The decomposition into compatible and force balanced components of an imposed edge swelling $\mathbf{w}_0$, representing changes in the rest length. }
     \label{fig:edge_swelling_projection}
\end{figure}
               
\begin{figure}
    \centering
    \includegraphics[width=0.9\linewidth]{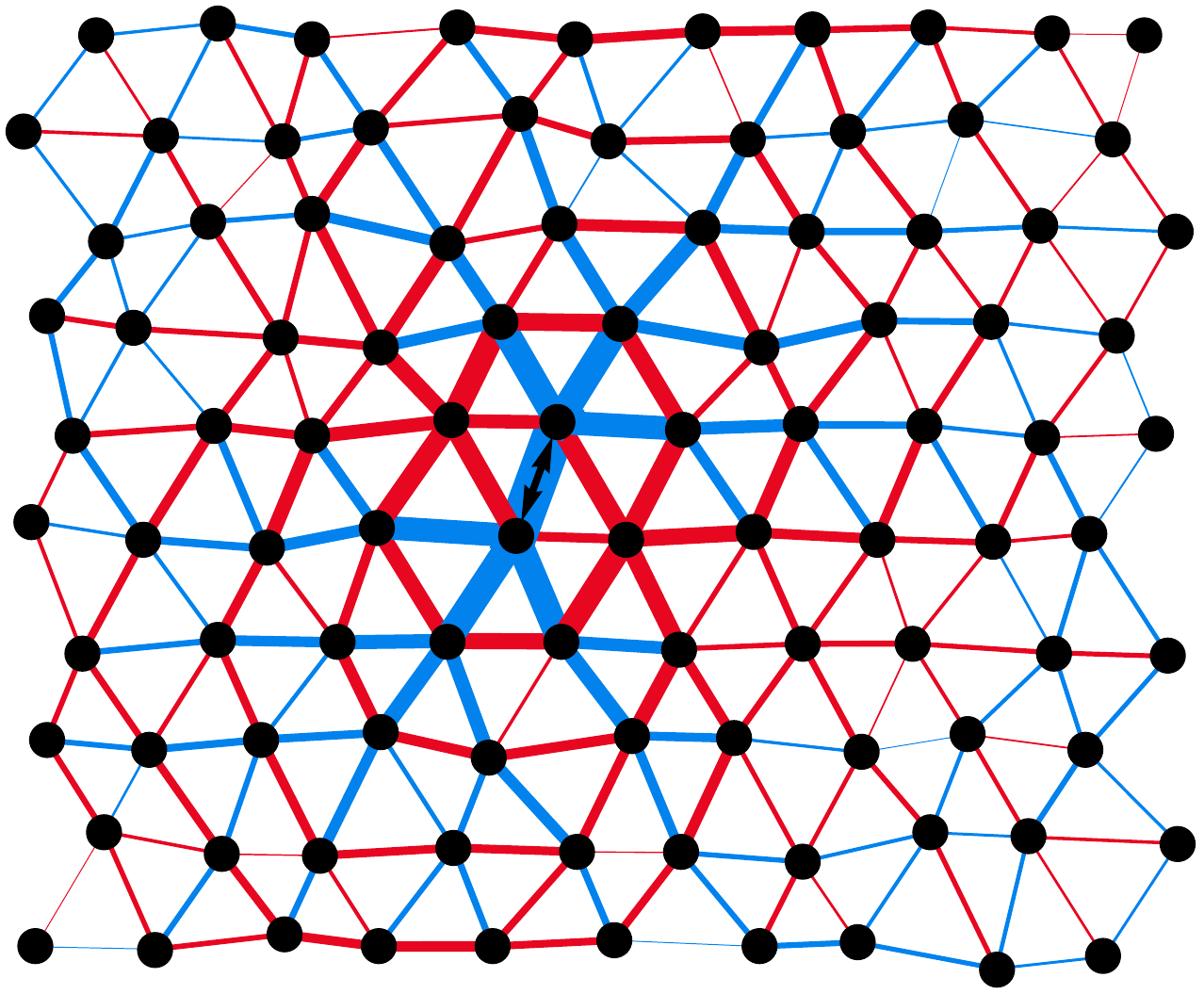}
    \caption{The rest length of an edge in a disordered triangular lattice is increased (black arrows).
    The system is incapable of reducing potential energy to zero in response to this rest-length change, so some edges carry tension (magnitude proportional to edge thickness) from being compressed (blue) or extended (red).}
    \label{fig:responsetoedge}
\end{figure}

Note the vector $\mathbf{w}_0$ can be decomposed as shown in Fig.~(\ref{fig:edge_swelling_projection}).   Only the compatible component can be relaxed via node displacements.  Once all of this is relaxed only the force balanced component remains. Therefore, the minimal $\mathbf{w}$ can be obtained by projecting the imposed change into $\textrm{null}(Q')$. As we defined, columns of $S'$ form an orthonormal basis for $\textrm{null}(Q')$, then minimal $\mathbf{w}$ is given by
\begin{equation}
    \mathbf{w}=S'S'^T\mathbf{w}_0 .
\end{equation}
In terms of elongations and tensions this is:
\begin{equation}
    \mathbf{t}=\sqrt{K}S'S'^T\sqrt{K}\mathbf{e}_0 ,\label{eq:edge_swelling_nice}
\end{equation}
where $\mathbf{e}_0$ are the imposed elongations.  
An example of the computation of tensions in response to edge swelling is shown in Fig.~(\ref{fig:responsetoedge}).
A set of displacements that minimize $|\mathbf{w}|$ can be obtained by use of the pseudo inverse: 
\begin{equation}
    \mathbf{U}=\left[C'\right]^+ \mathbf{w}_0 .
\end{equation}
The set of displacements  obtained via pseudo inverse will be orthogonal to all zero modes. 

It is worth noting how Eq.~(\ref{eq:edge_swelling_nice}) looks when we don't do the change of variables $C \rightarrow C'$.  Let $S$ be a matrix such that its rows are an orthonormal basis for $\text{null}(Q)$.  Note that for one dimensional mechanical networks, and scalar flow problems $\textrm{null}(Q)$ is just the cycle space. Then one can show,
\begin{equation}
    \mathbf{t}=S^T \left[S K^{-1} S^T \right]^{-1}S\mathbf{e}_0.
    \label{eq:edge_general}
\end{equation}
This is a more complicated expression than Eq.~\eqref{eq:edge_swelling_nice}  but it is also a more general expression since it does not depend on $K$ being a positive diagonal matrix.  We will see that analogous expressions are applicable to the case of prestressed networks and dynamics. In all of these cases the tension, or its analogue, only depends on the incompatible part of the applied edge elongations, or its analogues, a consequence of Kirchhoff's laws. 
Similar results have showed up in different contexts in Ref.~\cite{lubensky2015phonons,mao2018maxwell,prestressed_elasticity}.
For one dimensional spring networks this $S$ is an orthonormal basis for the cycle space. 
       
\subsection{Response to prescribed displacements on nodes}
Consider a spring mass network where some agent selects a set of nodes---let us label the set as A and all other nodes as B---and produces displacements $\mathbf{u}_A$ on them without directly interacting with any other nodes. The response of the network is obtained by solving the minimization problem:
\begin{equation}
    \underset{\mathbf{U}_B}{\text{min}} \;  |\mathbf{w}|^2 \quad \text{subject to } \; \mathbf{w}=C'_A\mathbf{U}_A + C'_B \mathbf{U}_B.
\end{equation}
Here we have split the $C'$ matrix in two. $C'_A$ contains only the columns identified with nodes in $A$ while the rest of the columns are in $C'_B$.  This problem is very similar in form to the edge swelling problem of the previous section in Eq.~\eqref{edge_swelling_problem}.
We therefore solve it analogously:
\begin{equation}
    \mathbf{U}_B=-[C'_B]^+ C'_A\mathbf{U}_A,
\end{equation}
\begin{equation}
    \mathbf{w}_{\text{min}}=  S_BS_B^T \,  C'_A\mathbf{U}_A ,
\end{equation}
where $S_B$ is an orthonormal basis for $\text{null}(Q_B)$ and $Q_B=C_B^T$. 
Therefore, the solution for $\mathbf{w}_{\text{min}}$ will produce no net forces on the nodes of set $B$, which is an obvious physical requirement for the static response of any system.
            
\begin{figure}
    \centering
    \includegraphics[width=0.9\linewidth]{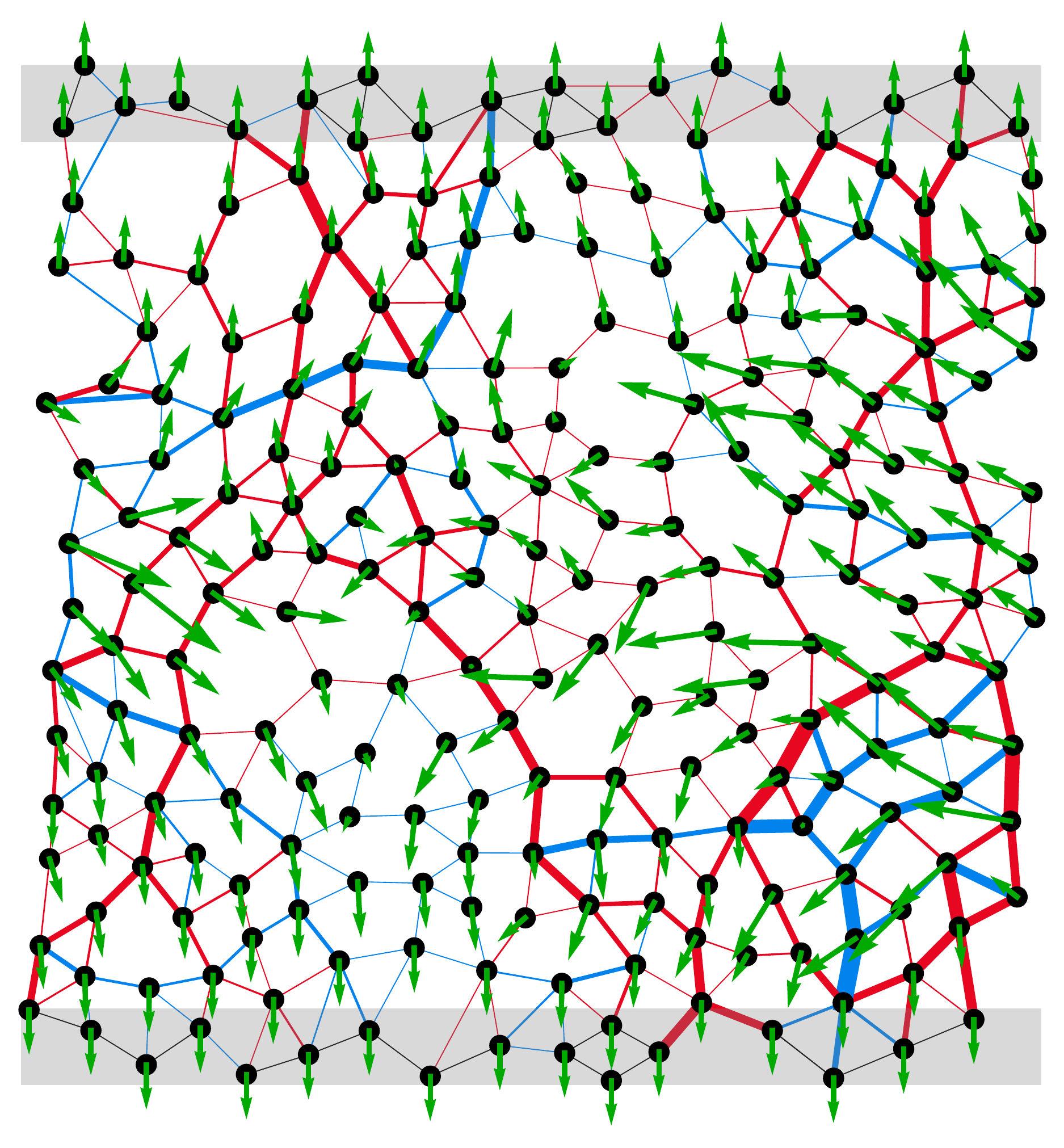}
    \caption{The response of a mechanical network to prescribed displacements (nodes within shaded gray regions). Displacements of all nodes shown by green arrows. 
    The resulting tensions (magnitude proportional to edge thickness) are colored red (blue) if extended (contracted).}
    \label{fig:responsetoprescribeddisplacement}
\end{figure}

We can alternatively solve the problem with the dynamical matrix.  We first split said dynamical matrix into blocks for the subspaces $A, B$:
\begin{equation}
    -\begin{pmatrix} \mathbf{F}_{A} \\ 
    \mathbf{0}\end{pmatrix} =\begin{pmatrix} D_{AA} & D_{AB} \\ 
    D_{BA} & D_{BB} \end{pmatrix} \begin{pmatrix} \mathbf{U}_{A} \\  
    \mathbf{U}_{B} \end{pmatrix}.
    \label{eq:pdisplacements}
\end{equation}
Here, $D_{BA}$ is the submatrix made up of the rows that map to nodes in $B$ and the columns that map to nodes in $A$, the other blocks are similarly defined.  To solve this problem we first look for the displacements of B by solving the bottom row,
\begin{equation}
    -D_{BA}\mathbf{U}_{A}=D_{BB}\mathbf{U}_{B}  .\label{eq:solveforB}
\end{equation} 
Once $\mathbf{U}_{B}$ is known, we can solve for $\mathbf{F}_{A}$ using the top row of Eq.~\eqref{eq:pdisplacements}. An example of the computation of tensions and displacements in response to applied displacements is shown in Fig.~(\ref{fig:responsetoprescribeddisplacement}).
The solution for the force is unique, but the solution for the displacements in $B$ may not be unique due to zero modes. 
These modes, by definition, do not alter the energy cost of a deformation. 
This type of calculation, where the dynamical matrix is separated into controlled and free nodes, has appeared in Ref.~\cite{rocklin2021elasticity}.
            
\subsection{Condensing nodes}

There are a number of situations in which one would wish to constrain a group of nodes to move with the same displacement, or in other words have the same potential. This can be useful for cases such as defining coarse-grained degrees of freedom which represent a set of nodes in a network. Note that we call this procedure ``condensing'' the nodes, but we are not shrinking the physical distances between them.  We simply require that they move together as a rigid body. 

The compatibility matrix has one column for each degree of freedom. Condensing a group of nodes into a super node supposes diminishing the degrees of freedom and so we must change the compatibility matrix. The super node will have $d$ columns in the new compatibility matrix. The column for super node displacement along the $x$ direction is the sum of the $x$ direction columns of each of the nodes condensed into the super nodes, representing a combined constraint on the super node from all the individual constraints in the original network. We do the same procedure for all the spatial directions.

\begin{figure}[h]
    \centering
    \includegraphics[width=0.45\textwidth]{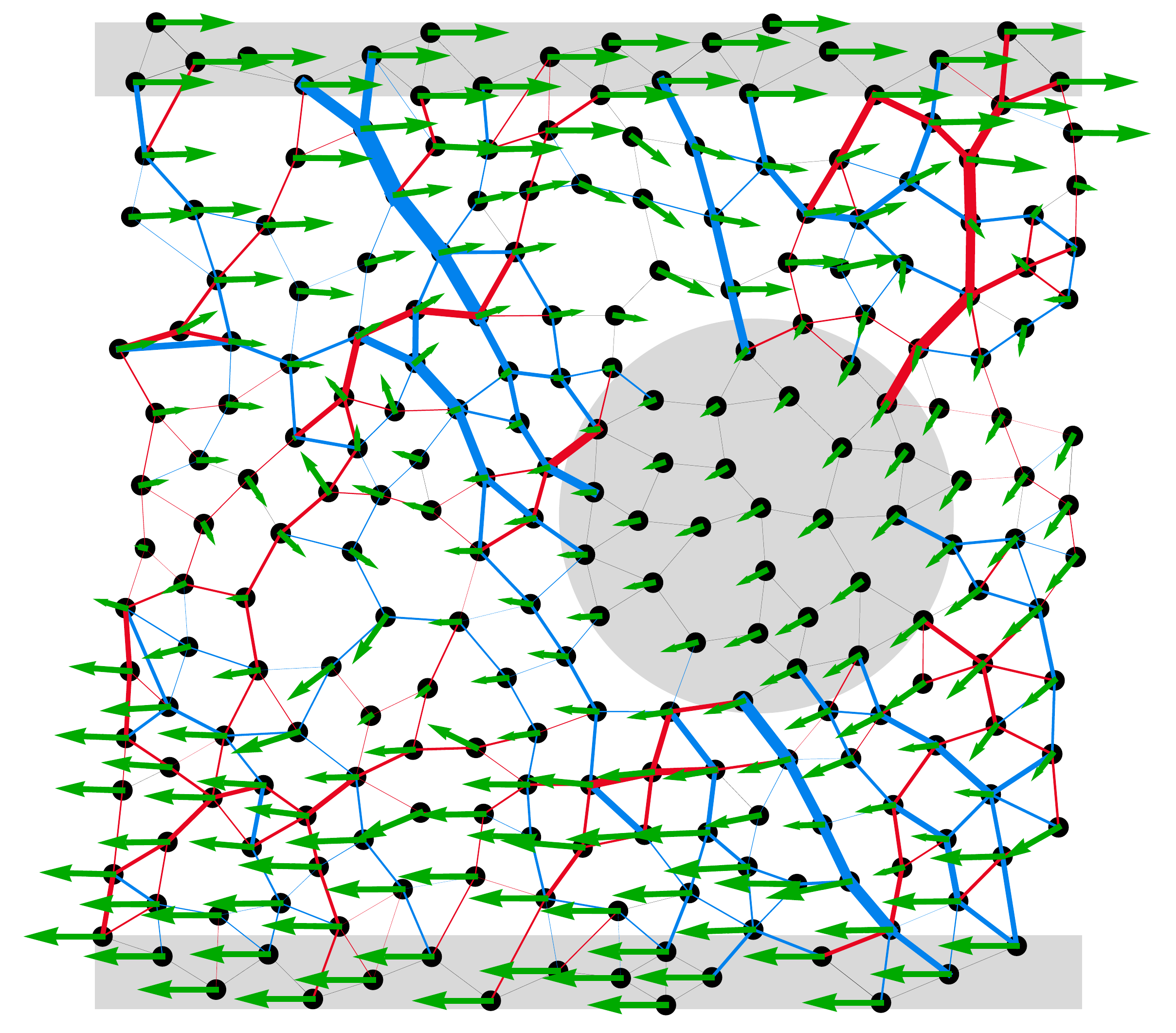}
    \caption{The response node displacements (arrows) and edge tensions (proportional to edge thickness, blue for compressed, red for extended) of a mechanical network with a number of nodes condensed into a rigid body (nodes within the gray circle) under shear (imposed displacements within shaded rectangles).
    As a rigid body, the displacements of the nodes within the gray circle must be a combination of an overall $x, y$ translation and $\theta$ rotation.
    Edges between two nodes in the rigid body cannot carry tension, as the distance between the two nodes they connect does not change, thus each node within the rigid body is not necessarily force-balanced from the edges alone.
    However, the rigid body can ``transmit'' tension across itself and must be force-balanced as a whole.
    }
    \label{fig:condense}
\end{figure}

More generally one can constrain the relative motion of a subset of nodes to be within a prescribed linear subspace. One natural example in mechanics would be to make the group move together as a rigid body, allowing rigid rotations. We will now describe the method for two dimensions. Let the $3$ degrees of freedom of a super node be $V_\alpha,V_\theta$ , where $\alpha$ goes over $x,y$ directions and $V_\theta$ is the rotation in radians. The displacement of a node included in this super node is therefore:
\begin{equation}
    \mathbf{U}_i=\mathbf{V}+V_\theta \, \hat{z}\times\left(\mathbf{X}_i-\mathbf{X}_0 \right),
\end{equation}
where $\mathbf{X}_i$ is the position of the node and $\mathbf{X}_0$ is the center of rotation which can be arbitrarily chosen. In index notation, this is:
\begin{equation}
    U_{i\alpha}=V_\alpha+V_\theta \, \epsilon_{\alpha\beta}\left(X_i^\beta - X_0^\beta\right).
\end{equation}
Now we can define a matrix $\Gamma$ that maps the degrees of freedom of the rigid cluster ($\mathbf{V}$) to the displacements of individual nodes ($\mathbf{U}$):
\begin{equation}
    \Gamma_{i\alpha}^{\mu}=\delta_{\alpha}^{\mu}+ \delta_{\theta}^{\mu}\epsilon_{\alpha\beta}\left(X_i^\beta - X_0^\beta\right),
\end{equation}
such that:
\begin{equation}
    U_{i\alpha}= \Gamma_{i\alpha}^{\mu}V_\mu.
\end{equation}
Note that the elements $V_\mu$ are $V_x,V_y,V_\theta$. 

Now we group together all the columns of the original compatibility matrix that map to nodes in the super node into a matrix we call $ C_{A}$, where the rest of the columns we group into $ C_{B}$, the elongations are given by:
\begin{equation}
    \mathbf{e}= \begin{pmatrix}  C_{B} ,   C_{A} \Gamma  \end{pmatrix} \begin{pmatrix} \mathbf{U}_{B}\\\mathbf{V} \end{pmatrix}.
\end{equation}
By grouping the two matrices $C_B, C_A \Gamma$ together, we have formed a new matrix that plays the role of the usual compatibility matrix from before, but that now acts on the degrees of freedom of the rigid cluster instead of the degrees of freedom of its components.
This ``new'' compatibility matrix's transpose, the new equilibrium matrix, obeys a similar expected relationship:
\begin{equation}
    \begin{pmatrix} \mathbf{F}_{B}\\\mathbf{F}_V
    \end{pmatrix}=\begin{pmatrix} Q_{B} \\  \Gamma^T  Q_{A}\end{pmatrix}\mathbf{t}.
\end{equation}
Note that $\mathbf{F}_V$ has 3 components that correspond to the forces along each degree of freedom of the super node. In this case the rotational component would be the net torque on the super node around $\mathbf{X}_0$. 
The procedure described here seems quite generic, but this is simply because the information regarding how nodes are constrained is encapsulated entirely within the matrix $\Gamma$. 
With any matrix $\Gamma$ that is suitable for a given system, one can solve  linear response problems of the previous section on that system by simply using the new compatibility and equilibrium matrices in place of the ``old'' compatibility and equilibrium matrices used previously.
An example of the computation of tensions and displacements in response to applied displacements in a system with condensed nodes is shown in Fig.~(\ref{fig:condense}).

One common scenario where nodes are condensed is to make a ground node. In mechanics, the resulting network is called a pinned graph, and the grounded nodes are called pins (since they are ``pinned'' to the ground). To ground a group of nodes we follow the general procedure described above but make $\Gamma=0$. The new compatibility matrix is simply $C_B$ .  

\subsection{Response to global strains and stresses}
Most of the preceding sections offer a microscopic perspective. By this we mean that the forces and displacements are specified at the level of individual nodes or edges. Materials science is most often concerned with macroscopic quantities, stresses strain and elastic moduli. In this section we describe the response of a network to macroscopic stresses and strains and compute elastic moduli in two approaches: a node-based approach through the control of boundary nodes, and an edge-based approach where a macroscopic strain is treated as analogous to changes in rest lengths.

\subsubsection{Node-based approach}
Consider placing a spring network, 2D for this example, inside a square box. We fix some of its nodes to the boundary box, creating a super node with the method described in the previous section (Fig.~\ref{fig:sameTensions}, a).  We then allow this boundary to deform as prescribed by an affine strain $\epsilon$.  In 2D the strain tensor has three independent components $(\epsilon_{xx},\epsilon_{yy},\epsilon_{xy})$ therefore we have reduced the degrees of freedom of the boundary nodes to three. We can find a matrix $\Gamma$ such that:
\begin{equation}
    \mathbf{U}_{A}=\Gamma\boldsymbol{\epsilon} ,
\end{equation}
where we represent the strain as a $3\times1$ vector, known as the Voigt notation~\cite{voigt1910lehrbuch}. Here $\mathbf{u}_{A}$ is the displacements of the nodes in the boundary, where external forces are applied. As in the previous section, we arrive at equation:
\begin{equation}
    \mathbf{e}=  C_{B}\mathbf{U}_{B} +  C_{A}\Gamma\boldsymbol{\epsilon}, \label{eq:BAuepsilon}
\end{equation}
where $C_B$ and $\mathbf{U}_B$ are the compatibility matrix and the displacements of the nodes in the bulk of the material respectively. We then have a new compatibility matrix:
\begin{equation}
    C=\begin{pmatrix}  C_{B} ,   C_{A} \Gamma  \end{pmatrix},
    \label{eq:globalCompMat}
\end{equation}
which also gives us a new $Q=C^T$  and the following relationships:
\begin{equation}
    \mathbf{e}=C\begin{pmatrix}
    \mathbf{U}_B \\ \epsilon_{xx} \\ \epsilon_{yy} \\ \epsilon_{xy} 
    \end{pmatrix}, 
    \label{eq:strain_C}
\end{equation}
\begin{equation}
    \begin{pmatrix}
    \mathbf{F}_B \\\sigma_{xx} \\ \sigma_{yy} \\ \sigma_{xy}
    \end{pmatrix}=Q\mathbf{t}. 
    \label{eq:stress_Q}
\end{equation}
Here, $\mathbf{F}_B$ is the force by the network on the bulk nodes and the $\sigma_{\mu\nu}$ is the components of the stress tensor.  We can verify that these are indeed the components of the stress tensor by taking the derivative of the elastic energy with respect to the strain. This now allows us to do linear response problems where we prescribe a macroscopic strain or stress and find the edge extensions and node displacements in the bulk.  These problems are solved in the same way we solved the problems of prescribed displacements (now prescribed strains) and prescribed forces (now imposed stresses) (as shown in Fig.~\ref{fig:sameTensions}a).

\subsubsection{Edge-based approach}
\begin{figure}[h!]
    \centering
    \includegraphics[width=\linewidth]{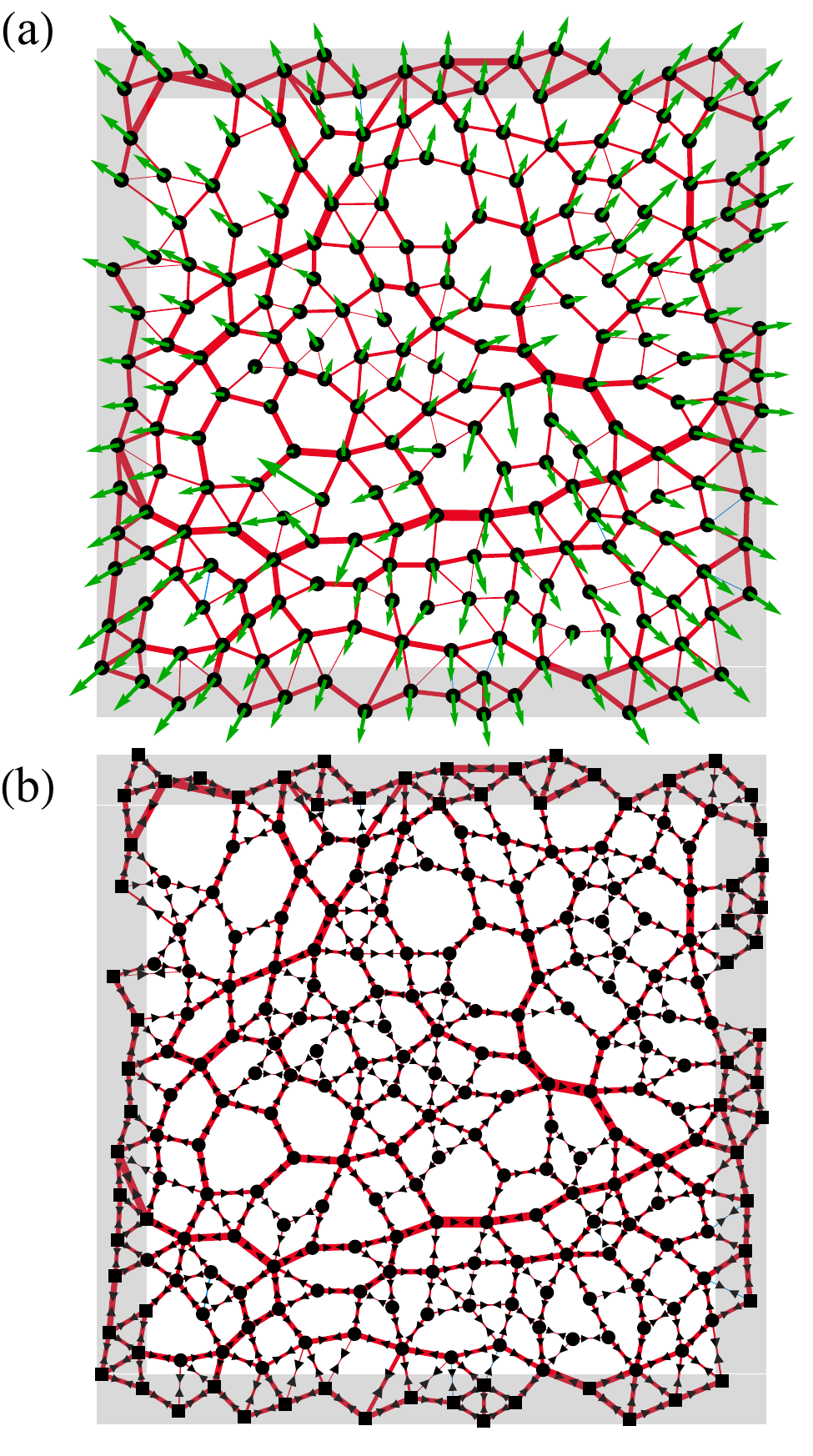}
    \caption{Response of a mechanical network to a global strain. 
    (a) The node-based approach to finding response to a global strain (here, global dilation), is accomplished by prescribing affine displacements to nodes on the boundary (within the shaded region).
    Resulting displacements for the bulk (arrows) and tensions (proportional to edge thickness) can be calculated using Eq.~(\ref{eq:solveforB}) with the new compatibility matrix from Eq.~(\ref{eq:globalCompMat}).
    (b) The edge based analog to global strain requires imposing (negative) edge elongations according to a given affine strain (again, contraction; arrows on edges, from Eq.~(\ref{eq:globalStrainElongations}), while pinning the boundary (squares, within shaded region).
    The resulting tensions (proportional to edge thickness, red for extension) are identical to the node based approach if both systems are under the same global strain.             }
    \label{fig:sameTensions}
\end{figure}

We have just described a node-based approach to prescribing a global affine strain. Now we  describe an edge-based approach. The two methods are equivalent for the purposes of calculating elastic moduli under ``fixed" or ``hard wall" boundary conditions but the edge-based method can be applied in more scenarios such as periodic boundary conditions where there is no obvious choice of ``boundary nodes."
Two examples, one of the node-based approach, one of the edge-based approach under hard wall boundary conditions, are shown in Fig.~(\ref{fig:sameTensions}).
Note that the tension response of these two examples is exactly equal.

\begin{figure}[h]
    \centering
    \includegraphics[width=0.45\textwidth]{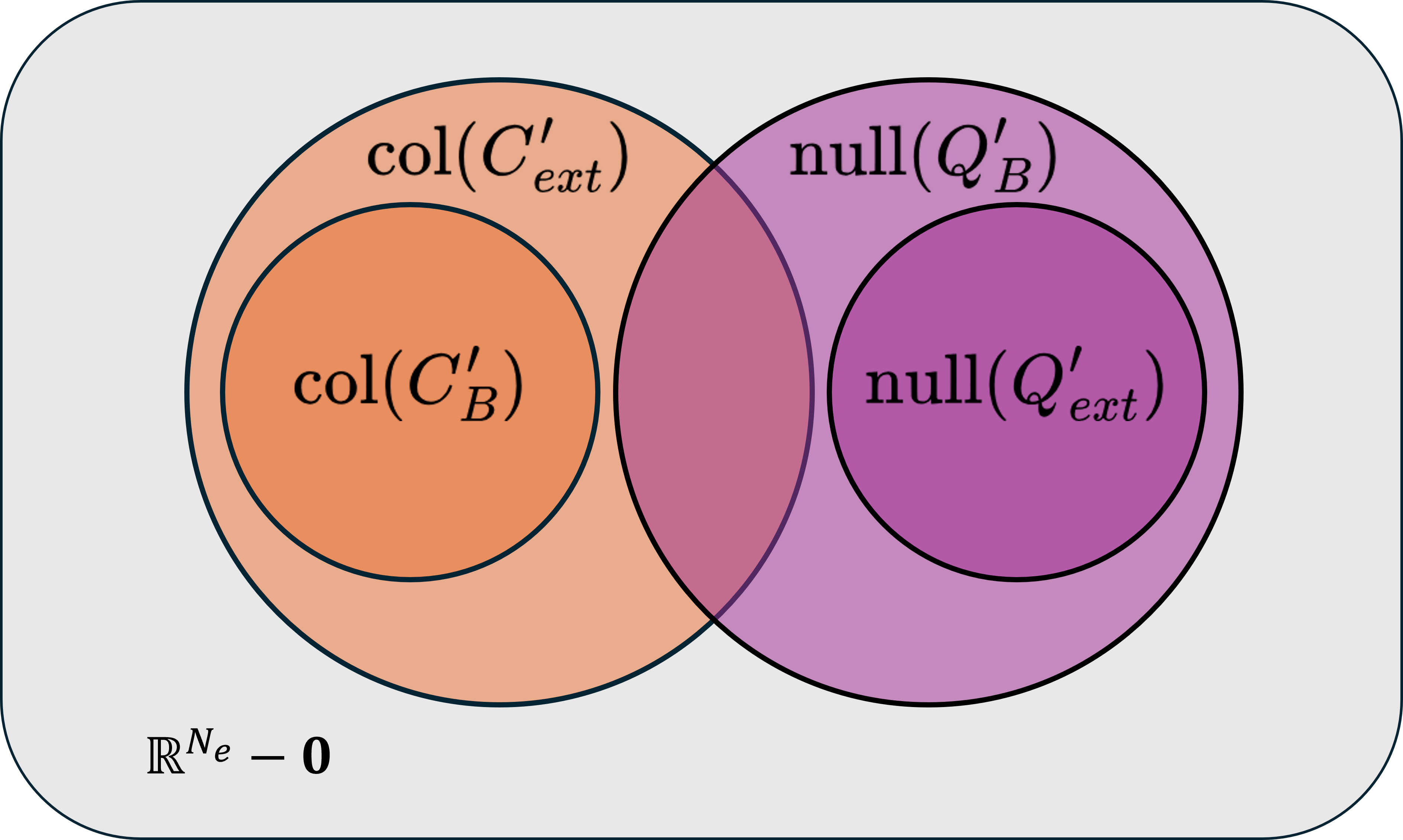}
    \caption{Venn diagram representation including all vectors in $\mathbb{R}^{N_e}$ with the exception of the zero vector $\mathbf{0}$. We show here the relationship between the relevant linear subspaces.   Note that the intersection corresponds to responses of the bulk to forces or displacements of the boundary. }
    \label{fig:venncompstress}
\end{figure}

\begin{figure}
    \centering
    \includegraphics[width=\linewidth]{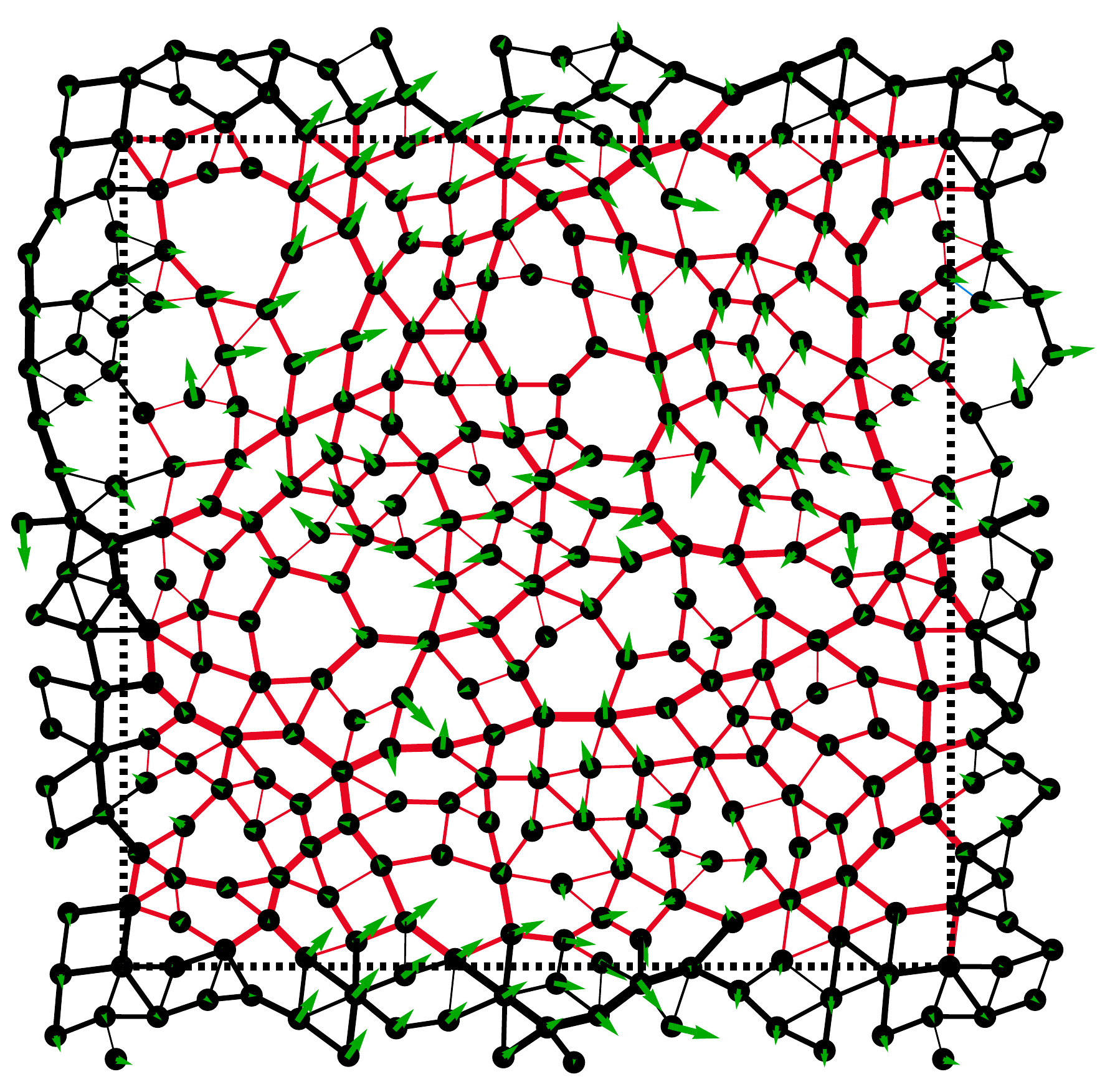}
    \caption{A mechanical network subject to uniform edge rest length decreases under periodic boundary conditions (with the boundary denoted by dashed lines).
    Edge tensions (proportional to edge thickness) show that almost all edges are extended (red). Node displacements are denoted by green arrows. 
    }
    \label{fig:periodicBoundary}
\end{figure}

\begin{figure}
    \centering
    \includegraphics[width=\linewidth]{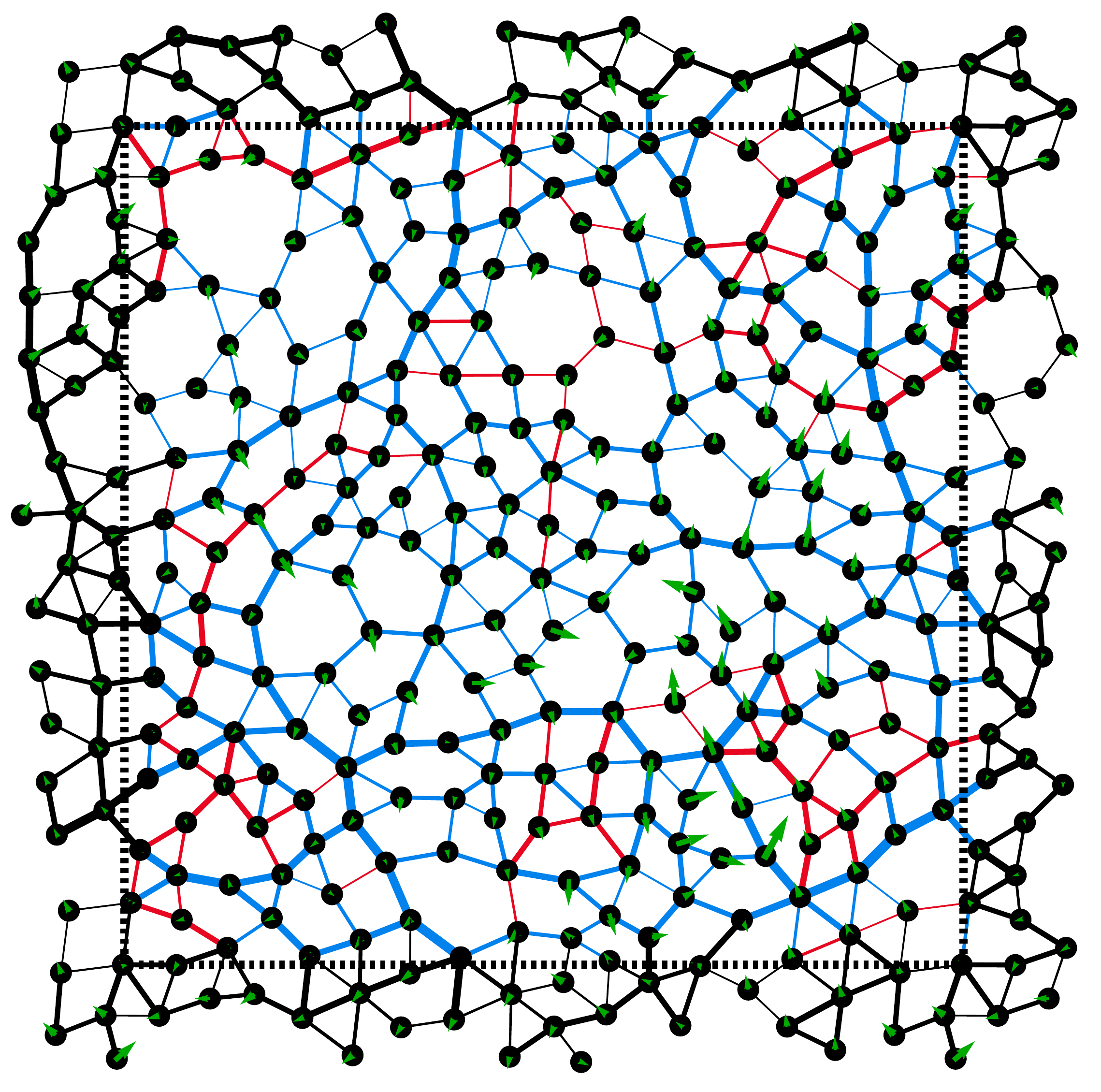}
    \caption{A mechanical network subject to pure shear (via edge elongations) under periodic boundary conditions (with the boundary denoted by dashed lines).
    Edge tensions (proportional to edge thickness) show that some edges are compressed (blue) and extended (red), resulting in the displacement of nodes (green arrows).
    This response if found via Eq.~(\ref{eq:globalStrainElongations}), where the strain tensor is anti-diagonal with $\epsilon_{12}=\epsilon_{21}=-0.01$.
    }
    \label{fig:strain}
\end{figure}

The edge based approach is mathematically very similar to the edge swelling problem described earlier Eq.~\eqref{edge_swelling_problem}, where the imposed elongations are:
\begin{equation}
    e_{0,ij}= C_{\textrm{macro},ij}^{\mu\nu}\epsilon_{\mu\nu}.
    \label{eq:globalStrainElongations}
\end{equation}
This is the same as the ``affine extension'' ($e_{\textrm{aff}}$) used in Refs.~\cite{lubensky2015phonons,mao2018maxwell}. It is worth noting that although mathematically we treat this problem using the edge swelling formulation, the physical meaning is different.
In the edge swelling problem, we consider the change of the rest length of some edges due to internal mechanisms in the edge, whereas here we consider externally imposed macroscopic strain.  Following the methods in Refs.~\cite{lubensky2015phonons,mao2018maxwell}, we first find the ``affine extension'' using Eq.~\eqref{eq:globalStrainElongations}, and then project it to the space of states of self stress (obtained under appropriate boundary conditions), as we discuss below.

The matrix  $C_{\textrm{macro}}$ is obtained from the relationship between strains and elongations,
\begin{equation}
    C_{\textrm{macro},ij}^{\mu\nu}= \ell_{ij}\hat{\ell}_{ij}^\mu\hat{\ell}_{ij}^\nu 
\end{equation}
The vectors $\ell_{ij}, \hat{\ell}_{ij}^\mu$  are respectively the magnitude and direction of all edges $\ell_{ij}^\alpha=X_i^\alpha-X_j^\alpha$. Here the pair of Greek indices act as one index in Voigt notation. 
As defined above the matrix has four columns (in 2D) one for every component of the strain tensor $xx,yy,xy,yx$. Since the strain tensor is symmetric the $xy$ and $yx$ can be replaced with a single column which is the addition of the two. The resulting $C_{\textrm{macro}}$ will then have three columns and as many rows as there are edges. Keeping the four columns introduces a zero mode which corresponds to a global rotation.    We can obtain an extended compatibility matrix by concatenation $C_{\text{ext}}=[C , C_{\textrm{macro}}]$ such that
\begin{equation}
    \mathbf{e}=C_{\text{ext}}\begin{pmatrix} \mathbf{U} \\  \epsilon_{xx} \\\epsilon_{yy}\\ \epsilon_{xy}\end{pmatrix}.
\end{equation}
Fig.~(\ref{fig:venncompstress}) gives a diagramatic representation of $C_{\text{ext}}$, $C_B$, and their transposes.

For an imposed strain it does not matter if we used the node method or the edge method, both problems reduce to the edge swelling problem:
\begin{equation}
    \mathbf{e}=  C_{B}\mathbf{U}_{B} +\mathbf{e}_{0}.
\end{equation}
For the edge method, $\mathbf{e}_0=C_{\text{macro}}\boldsymbol{\epsilon}$, while for the node method, $\mathbf{e}_0=  C_{A}\Gamma\boldsymbol{\epsilon}$.
These two vectors are not the same in general. When we are dealing with fixed boundary conditions, their difference is compatible:
\begin{equation}
    C^{\textrm{macro}}\boldsymbol{\epsilon}-  C_{A}\Gamma\boldsymbol{\epsilon} \in \text{col}(C_B).
\end{equation}
This is due to the fact that both operators move the nodes in the boundary in the same way, according to strain $\epsilon$, but $C_{A}\Gamma\boldsymbol{\epsilon}$ does not result in bulk displacements. This means that on can go from one to the other via displacements of the nodes in the bulk.  

Note that the notion is ``fixed boundary condition'' is associated with the edge swelling formulation we use here. Translating to the problem of imposed strain, this correspond to rigidly deforming the boundary, which is the same as in the node-based approach.  Another way, as we mentioned above, is to use periodic boundary conditions across the box.  In this case, the imposed elongations are the same as given in Eq.~\eqref{eq:globalStrainElongations}, but the states of self stress are slightly different (due to different $C_B$), which will cause small differences in the response, but these differences will in general vanish for large systems.
Example of the computation of tensions and displacements in response to global strains through edge elongations are shown in Figs.~(\ref{fig:periodicBoundary}, \ref{fig:strain}).

The solution to the edge swelling problem tells us that,
\begin{equation}                    
    \mathbf{t}=\sqrt{K}SS^T\sqrt{K}\mathbf{e}_{0}, 
    \label{eq:te_edgeswelling}
\end{equation}
where the columns of $S$ are an orthonormal basis for the states of self stress, the null space of $Q_B'=Q_B\sqrt{K}$. The node and edge methods are equivalent since 
\begin{equation}
    \sqrt{K} C_{\text{bdry}}\boldsymbol{\epsilon}- \sqrt{K} C_{A}\Gamma\boldsymbol{\epsilon} \in \text{col}(\sqrt{K}C_B).
\end{equation}
This difference would then be annihilated by $S^T$. 

\subsubsection{Finding the elastic moduli}
As discussed above, given an imposed strain $\epsilon$, we can compute an effective stress tensor $\sigma$ as a response. We can then determine the effective stiffness tensor $\mathbb{K}$,        
\begin{equation}
    \sigma_{\mu\nu} = \mathbb{K}^{\alpha\beta}_{\mu\nu} \epsilon_{\alpha \beta}. \label{eq:effective_stiffness_tensor_def}
\end{equation}
Note, from Eq.~(\ref{eq:stress_Q}):
\begin{equation}
    \begin{pmatrix}
    \sigma_{xx} \\ \sigma_{yy} \\ \sigma_{xy}
    \end{pmatrix}=Q_A\mathbf{t}, 
\end{equation}
and 
\begin{equation}
    S^T\sqrt{K} C_{A}\Gamma\boldsymbol{\epsilon} =S^T \sqrt{K}\mathbf{e}_0.
\end{equation}
By substituting these expressions into Eq.~(\ref{eq:te_edgeswelling}), we obtain a linear relationship between stress and strain as in Eq.~(\ref{eq:effective_stiffness_tensor_def}) and identify the stiffness tensor as: 
\begin{equation}
    \mathbb{K}= \Gamma^T Q_A \sqrt{K}SS^T\sqrt{K} C_A  \Gamma.
\end{equation}

\subsection{Prestressed networks}\label{SEC:prestress}
Previously, we assumed that the equilibrium configuration we find the system in is stress free. Here we consider spring networks whose initial configuration is stressed and in equilibrium. The solutions of linear response problems will still minimize the elastic energy, but representing them as minimization problems in the same way as before is not as straightforward. 
We will discuss this in the context of two dimensional networks. The procedure is easily generalized to networks of higher dimensions by adding more $\perp$ components in a similar way the one that we will describe.

It has been shown in Ref.~\cite{prestressed_elasticity} that in two dimensions the change in elastic energy due to a small node displacements $\mathbf{U}$ around equilibrium is given by:
\begin{equation}
    \delta E=\frac{1}{2}\mathbf{U} \begin{pmatrix}
    Q^\parallel & Q^{\perp}
    \end{pmatrix}
    \begin{pmatrix}
    K^{\parallel} & 0 \\ 0 & K^{\perp} 
    \end{pmatrix} 
    \begin{pmatrix}
    C^{\parallel} \\ C^\perp
    \end{pmatrix}\mathbf{U},
\end{equation}  
where $Q^{\parallel},C^{\parallel} $ are the usual equilibrium and compatibility matrix and $Q^{\perp},C^{\perp}$ are the equilibrium and compatibility matrix of the network if it where rotated by $90^{\circ}$. We previously defined $C^\perp$ in  Sec.~\ref{SEC:physical}. The $K^{\parallel}$ matrix is equal to the usual diagonal matrix of spring stiffness (as in Eq.~(\ref{eq:dmatDef})), while 
\begin{equation}
    K_{ij}^{\perp nm}=  -\frac{t_{ij}}{\ell_{0,ij}}\delta_{ij}^{nm},
\end{equation}
coming from the second order expansion of Eq.~\eqref{EQ:nonlineart}, as shown in Ref.~\cite{prestressed_elasticity}.  
Here, $t_{ij}$ is the tension of edge $ij$ at equilibrium and $\ell_{0,ij}$ is the rest length of the spring. We consider $t_{ij}$ to be positive when it pushes the nodes apart and negative when it pulls them together.  

One can then represent edge elongations as:
\begin{equation}
    \mathbf{e}=\begin{pmatrix}\mathbf{e}^{\parallel} \\ \mathbf{e}^{\perp}\end{pmatrix}= \begin{pmatrix}
    C^{\parallel} \\ C^\perp
    \end{pmatrix}\mathbf{U}. 
\end{equation}
Note, now we have a vector for each edge. The change in tension due to the displacements is given by:
\begin{equation}
    \delta\mathbf{t}=\begin{pmatrix} \delta\mathbf{t}^\parallel\\ \delta\mathbf{t}^{\perp}  \end{pmatrix} = \begin{pmatrix}
    K^\parallel C^{\parallel} \\ K^\perp C^\perp
    \end{pmatrix}\mathbf{U}. 
\end{equation}
    
We can define $Q=\begin{pmatrix}Q^\parallel & Q^{\perp}\end{pmatrix}$, $C=Q^T$ and $K=\begin{pmatrix}K^{\parallel} & 0 \\ 0 & K^{\perp} \end{pmatrix}$ and write equations analogous to the non prestressed case. Linear response problems here are better solved by use of the dynamical matrix $D=QKC$. We have previously considered the response  to perturbing rest lengths, in prestressed systems we can also perturb the stiffness and obtain non trivial response. We can differentiate Hooke's law to obtain: 
\begin{equation}
    \delta t^\parallel_{ij}= -k_{ij}e^{\parallel} -(\ell_{ij} -\ell_{0,ij})\delta k_{ij} + k_{ij}\delta \ell_{0,ij},
\end{equation}
\begin{equation}
    \delta t^{\perp}_{ij}= -\frac{t_{ij}}{\ell_{0,ij}} e^{\perp}_{ij}.
\end{equation}        
We can now solve the general linear response problem for prestressed networks with equation:
\begin{equation}
    0=-\begin{pmatrix}
    D,-QK^{\parallel}, Q\Lambda
    \end{pmatrix} \begin{pmatrix}
    \mathbf{U} \\ \delta\boldsymbol{\ell_0} \\ \delta \mathbf{k} 
    \end{pmatrix} + \mathbf{F}.
\end{equation}
where the $\Lambda$ is a diagonal matrix whose elements are $(\ell_{ij} -\ell_{0,ij})$. This problem is solved in an analogous fashion to the prescribed displacement problem. 

The solution to the edge swelling problem in prestressed networks is given by
\begin{equation}
    \delta\mathbf{t}=S^T \left[S K^{-1} S^T \right]^{-1}S\delta{\boldsymbol{\ell}}_0  
\end{equation}
which is essentially the same as Eq.~\eqref{eq:edge_general} where $S$ is such that its rows are an orthonormal basis for $Q$ as defined in this section and $K$ as it is defined in this section.
                
\section{Steady state response of irreversible transport problems }\label{SEC:irreversible}
\subsubsection{Entropy production in irreversible transport}
Many if not most transport phenomena are irreversible, meaning that they dissipate energy and produce entropy. This applies to electrical, diffusive and thermal transport to give a few examples.  We will see that we can treat these problems with the same tools we have developed in previous sections. We will describe the physics in terms of graph potentials and flows and reduce linear response problems to minimization problems solvable through the use of the fundamental theorem of linear algebra.

First, the minimized quantity will be the entropy production rate, often referred to as the entropy production. This has been called the law of minimum entropy production, which was introduced by Prigogine and stated as: 
``In the linear regime, the total entropy production rate in a system subject to flow of energy and matter, reaches a minimum value at the nonequilibrium stationary state"~\cite{Kondepudi_Prigogine_2015}.

Irreversible transport phenomena can be described through the language of thermodynamic forces and flows. This and the other relevant thermodynamic concepts in this section are explained in Ref.~\cite{Kondepudi_Prigogine_2015}. 
The entropy production per unit volume $\dot{\mathfrak{s}}$, is given by the sum of the products of thermodynamic forces which are gradients of potentials $\phi^{K}$ and flows which are current densities $\mathbf{J}^{K}$,
\begin{equation}
    \dot{\mathfrak{s}}=-\nabla \phi_{K}\cdot\mathbf{J}^{K}.\label{eq:epdensity}
\end{equation}
Above, the index $K$ is summed over and labels the different types of transport---electrical, thermal, diffusive, etc. Note that $\dot{\mathfrak{s}}$ is not the rate of change of the entropy but the entropy production rate, the difference is that the second quantity does not take into account exchange of entropy with the surroundings. When the system is close to equilibrium, in the ``linear regime" the forces have a linear relationship with the flows.:
\begin{equation}
    \mathbf{J}^{K}= -L^{J}_{K}\nabla \phi_J  .
    \label{eq:linear_constitutive}
\end{equation}

By the second law of thermodynamics, $\dot{s}\geq 0$, where the equality only being satisfied when all currents are zero. This implies that the matrix $L$ is positive definite.  Therefore, the matrix is invertible. It is also known that this matrix is symmetric---the Onsager reciprocal relations~\cite{PhysRev.37.405,PhysRev.38.2265}.

\begin{figure}[h]
    \centering
    \includegraphics[width=0.95\linewidth]{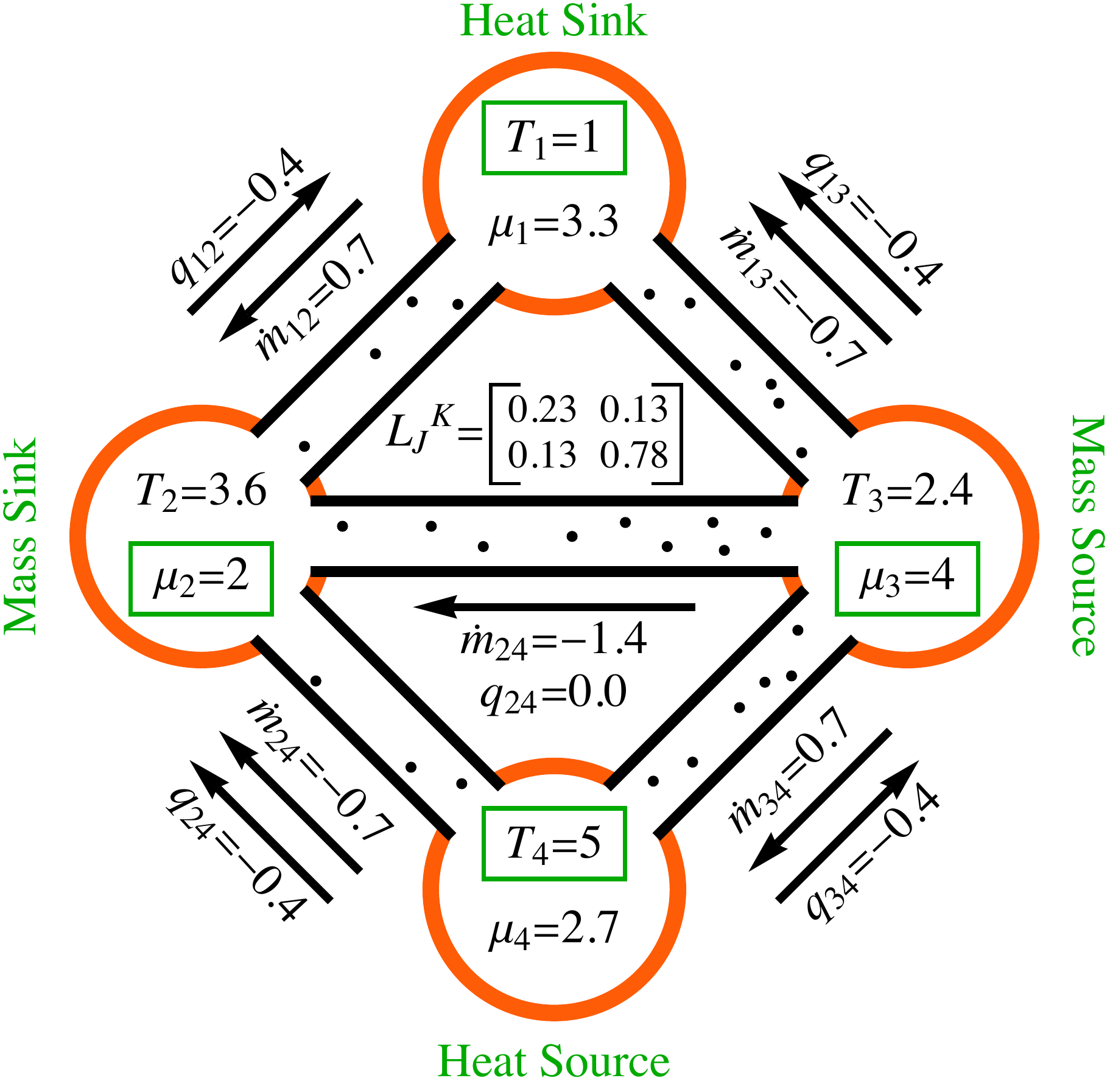}
    \caption{A four-node system where heat and mass are both able to flow between nodes.
    nodes 1, 4 are connected to a heat sink/source (fixed held temperatures in green, $T_1=1, T_4=5$) while nodes 2, 3 are connected to a mass sink/source (fixed held chemical potentials in green, $\mu_2=2, \mu_3=4$).
    The edges of this system are made from some hypothetical material where heat and mass transport are coupled such that the Onsager coefficients are: $L_{\text{heat}}^{\text{heat}} = 0.23$, $L_{\text{mass}}^{\text{heat}} = 0.13$, $L_{\text{mass}}^{\text{mass}} = 0.78$.
    The fact that the off-diagonal terms are non-zero ($L_{\text{mass}}^{\text{heat}} = 0.13$) indicates the coupling between heat and mass in this material.
    The calculated response of this system is shown by edge and node properties that are not in green, where arrows for edge properties point in the direction of positive flow.
    }
    \label{fig:coupled_transport}
\end{figure}

Now let us apply this to networks. Consider a system consisting of a network of pipes, wires or in general any conducting elements which come together at joints. We identify the joints with the nodes and the edges with the conducting elements. We make the assumption that the volume of the joints is negligible compared to the volume of the conducting elements. By integrating over volume, we can ``discretize" Eqs.~(\ref{eq:epdensity},\ref{eq:linear_constitutive}) on this network as: 
\begin{equation}
    \dot{\mathfrak{S}}=\left(\phi^m_K-\phi^n_K \right) \, i_{nm}^{K},
\end{equation}
\begin{equation}
    i_{nm}^{K}=\frac{a_{nm}}{\ell_{nm}} L^{K}_{J} \left(\phi_m^J-\phi_n^J \right),
\end{equation}
where $i^{K}_{nm}$ is the current of type $K$ going through conducting element $nm$ and $a_{nm},\ell_{nm}$ are the cross sectional area and the length of the conducting element respectively. We can define a generalized resistance matrix for each edge as:
\begin{equation}
    R_{nm}^{JK}=\frac{\ell_{nm}}{a_{nm}}\left[L^{-1}\right]^{JK}.
\end{equation}
Using this relationship, we can calculate the response on networks with coupled conductivity, a small example of which is shown in Fig.~(\ref{fig:coupled_transport}).

This resistance matrix is symmetric and positive definite, it inherits this quantities from $L$. To make the analogy with Ohm's law obviously we can label potential differences as ``voltages" ($v^{K}_{nm}=\phi^{K}_{n}-\phi^{K}_{m}$) to write the generalized Ohm's law:
\begin{equation}
      v_{nm}^{J}=-  R_{nm}^{JK} i_{nm,K}.
\end{equation}
Note the negative sign indicating that there is a voltage drop as we go from node $m$ to node $n$ given that the current goes from $m$ to $n$.  

The equation for the entropy production in terms of currents and resistances is now:
\begin{equation}
    \dot{\mathfrak{S}}=i^{nm,J} R^{K}_{nm,J}  i_{nm,K},
\end{equation}
Where $\dot{\mathfrak{S}}$ is the total entropy production rate. For electrical transport this reduces to the power dissipated $ri^2$. All of this terms are dissipative in nature. For this reason the ``law of minimum entropy production" can be understood as a law of least dissipation.

\subsection{Analogy between static mechanics and steady state in irreversible flow}

Here we will  write the equations for irreversible flow in the same form as the ones we used for mechanics. In this way we show that there is an analogy between mechanical equilibrium and thermal equilibrium. In their respective linear regimes we will show that entropy production plays the role of elastic energy. 

Currents are analogous to tensions, voltages to elongations and potentials to displacements, as we listed in Tab.~\ref{mapping}. The stiffness is analogous to conductance which is the inverse of resistance. We can write some basic relationships in terms of the incidence matrix $\mathcal{C}$ and its transpose $\mathcal{Q}$,
\begin{equation}
    v_{nm}^{K}=\mathcal{C}_{nm}^{p}\phi_{p}^K,
\end{equation}
\begin{equation}
    I_p^K=\mathcal{Q}_{p}^{nm} i_{nm}^{K},
\end{equation}
and the stiffness matrix:
\begin{equation}
    K_{pqK}^{nmJ}=\left[R_{pq}^{-1}\right]_{K}^J \delta_{pq}^{nm}.
\end{equation}
Note that this matrix is symmetric and positive definite. In mechanics the constitutive relation is what couples the different spatial components of potentials and flows, analogously here it is the constitutive relation that couples the different types of irreversible transport to each other.

We can also build a compatibility matrix:
\begin{equation}
    C_{pqK}^{nJ}=\mathcal{C}_{pq}^{n}\delta_{K}^J,
\end{equation}
where the equilibrium matrix would be its transpose. Now the total entropy production can be written as:
\begin{equation}
    \dot{\mathfrak{S}}=\boldsymbol{\phi}\,C^TKC\, \boldsymbol{\phi},
\end{equation}
which is of the same form as that of the elastic energy of a spring network. Note that the generalized dynamical matrix $D=C^TKC$ is semi positive definite. We can perform the same change of variables as in the mechanics case. Because $K$ is positive definite it can be decomposed as
\begin{equation}
    K=\sqrt{K}^T\sqrt{K}.
\end{equation}
We can include the conductances (``stiffnesses'') in the compatibility matrix by defining:
\begin{equation}
    C'=\sqrt{K}C,
\end{equation}
and describe currents and voltages with a single vector:
\begin{equation}
    \mathbf{w}=\sqrt{K}\mathbf{v} = \left(\sqrt{K}^T \right)^{-1}\mathbf{i}.
\end{equation}
The entropy production is then the norm squared of this vector:
\begin{equation}
   \dot{\mathfrak{S}}=|\mathbf{w}|^2.
\end{equation}
This allows us to solve linear response problems in a standardized manner just like it did for mechanical networks.\\

There are two basic response problems, imposing potential differences or imposing current sources. Current sources on the nodes are analogous to forces on the nodes on the mechanical case and imposing potentials is analogous to prescribing displacements. We have solved these problems in previous sections. These solutions apply identically to irreversible flow problems. 

Imposing potential differences on the edges is analogous to changing the rest lengths of the springs. For an electrical problem, rest-lengths are voltage sources, batteries \cite{frustrated_incompatible}. The analogy can be extended to voltage sources of other types, not just electrical. We are using the word voltage here to refer to any potential difference.  What these voltage sources are physically will depend on the context. 

In our search to find realizations of these voltage sources we might take a clue from the everyday battery, which is a chemical battery. Chemical reactions can be sources, or sinks, of heat, chemical species and charge. This naturally relates to thermal, diffusive and electrical transport respectively. In fact, chemical reactions  generally involve irreversibility and can be written in terms of thermodynamic forces and flows, included in Eq.~(\ref{eq:epdensity}) and coupled with the other irreversible processes via the matrix $L^{KJ}$. And so, it seems there is a great richness of phenomena within the linear regime that could be studied with the framework presented here.

For chemical reactions, the thermodynamic force is the affinity $A$, which depends linearly on the chemical potentials, but cannot generally be expressed as a potential difference. This is in contrast to transport phenomena. The thermodynamic flow is the velocity of reaction, also called ``rate of conversion,'' $\dot{\xi}$. The entropy production due to chemical reactions is: 
\begin{equation}
    \dot{\mathfrak{S}}_{chem}=\frac{A^{K}}{T}\dot{\xi}^K,
\end{equation}
where the index $K$ labels the different chemical reactions that may be taking place.

\section{Dynamics and waves}\label{SEC:waves}
In this section we present a more general analogy of dynamical processes on networks that involve both energy conserving, dissipative, and active components.  The mechanical equilibrium case presented in Sec.~\ref{SEC:static} and the thermal equilibrium case presented in Sec.~\ref{SEC:irreversible} can be seen as special cases of this more general analogy, where the simple frequency dependence of the edge and node admittance permits a neat scheme based on minimization.  Here, due to the more general dynamical rules, we will directly write dynamical equations on nodes and edges, instead of constructing a total energy or entropy.

\subsection{General analogy}
We start from the general physical network relations we wrote in Eq.~\eqref{EQ:electricall}. 
In the case of electric networks, $Y_n, y_{nm}$ represent node and edge admittance.  Using the frequency domain notation, familiar circuit elements have admittance
\begin{align}
    Y_{\textrm{capacitor}} &=i\omega C ,\\
    Y_{\textrm{resistor}} &= 1/R ,\\
    Y_{\textrm{inductor}} &= 1 / (i\omega L) .
\end{align}
The admittance of an edge or a node can be any combinations of these elements in parallel or series, which can be written as a complex number in general.

\begin{figure}[h]
    \centering
    \includegraphics[width=0.95\linewidth]{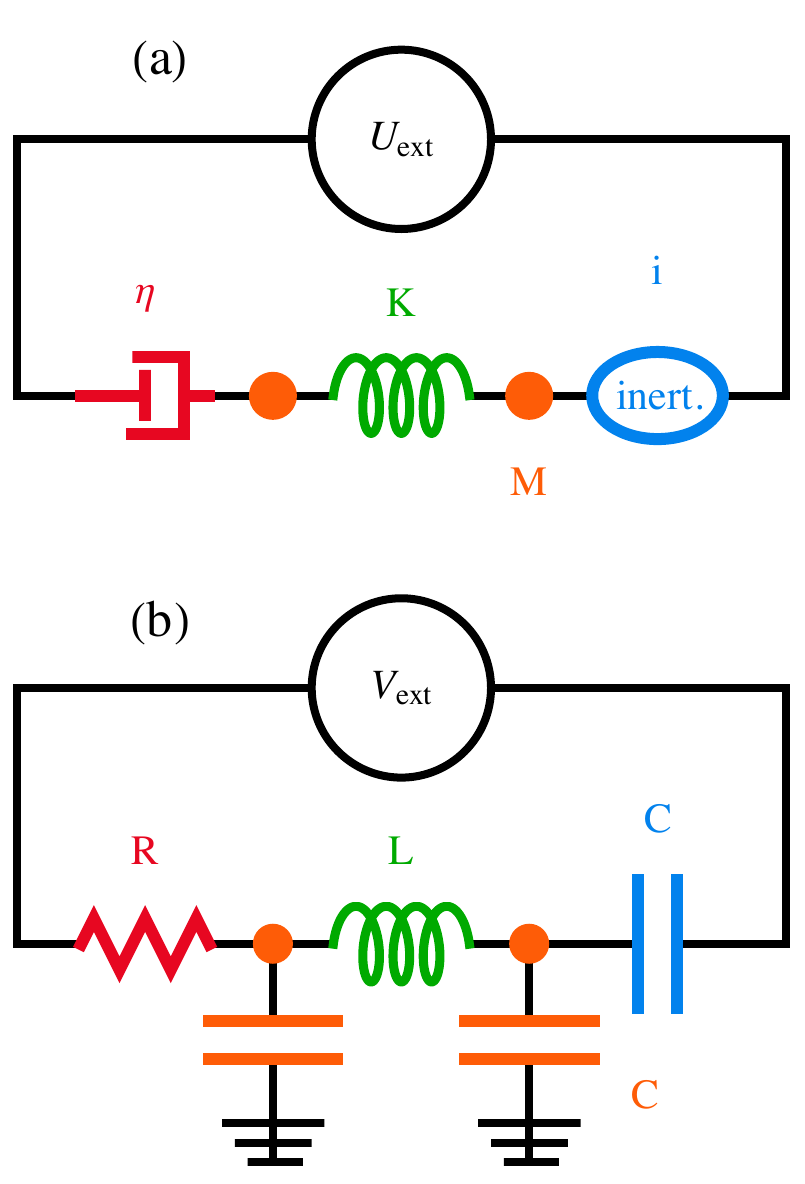}
    \caption{Analogous components (same color/position) between mechanical (a) and electrical (b) networks as described by Firestone's analogy~\cite{Firestone}. 
    }
    \label{fig:firestone}
\end{figure}

It is useful to write the mechanical problem in an analogous form, so we have a universal language for the dynamical response of networks.  
To do this, we adopt an analogy developed by Firestone in 1933 \cite{Firestone}.  Here, instead of node displacement, we take node velocity as potentials, while the flows remain forces.  
This allows a more natural analogy where the admittance of the mechanical problem correspond to, as shown in Fig.~(\ref{fig:firestone}):
\begin{align}
    Y_{\textrm{mass}} &=i\omega m ,\\
    Y_{\textrm{dashpot}} &= \eta \label{eq:dashpotY},\\
    Y_{\textrm{spring}} &= k / (i\omega) .
\end{align}

\noindent We describe other analogies between mechanical and electrical networks in the Appendix.

\subsection{Timescales of physical networks}
\begin{table*}[t]\label{mechanicalEOM}
\begin{tabular}
{|c|c|c|c|}
  \hline
  \diagbox{$\quad$nodes $Y$}{edges $y\quad$}\rule{0pt}{3ex}\rule[-2ex]{0pt}{0pt} & mass $i\omega m$ & dashpot $\eta$ & spring $\frac{k}{i\omega}$ \\
  \hline
  mass\rule{0pt}{3ex}\rule[-2ex]{0pt}{0pt} $i\omega M$ & $\delta=2\pi$ & $\omega''=\textrm{spec}(M^{-1}\mathcal{A})$ & $\quad\omega'=\sqrt{\textrm{spec}(M^{-1}\mathcal{K})}\quad$ \\
  \hline
  dashpot\rule{0pt}{3ex}\rule[-2ex]{0pt}{0pt} $H$ & $\omega''=\textrm{spec}(\mathcal{M}^{-1}H)$ & $\delta=\pi$ & $\omega''=\textrm{spec}(H^{-1}\mathcal{K})$ \\
  \hline
  spring\rule{0pt}{3ex}\rule[-2ex]{0pt}{0pt} $\frac{K}{i\omega}$ & $\quad\omega'=\sqrt{\textrm{spec}(\mathcal{M}^{-1}K)}\quad$ & $\quad\omega''=\textrm{spec}(K\mathcal{A}^{-1})\quad$  & $\delta=0$ \\
  \hline
\end{tabular}
\caption{Mechanical network dynamical timescales for different combinations of edge and node elements.}
\end{table*}

Beyond the static and steady state responses we discussed in Secs.~\ref{SEC:static},\ref{SEC:irreversible}, which are governed by the null space of the $\mathcal{Q,C}$ (or $Q,C$) matrices, we need to utilize the \emph{full spectrum} of these matrices to characterize the dynamical response, where forces don't balance on the nodes  and currents are not conserved, representing the conversion among different forms of energy.

Following the discussion in Sec.~\ref{SEC:physical}, the general equation of motion in the frequency domain, using matrix notation, takes the form
\begin{align}\label{EQ:EOMW}
    Y(\omega) V(\omega) = \mathcal{A}(\omega) V(\omega) + S(\omega) 
    +Q y(\omega)w, 
\end{align}
where 
\begin{equation}
    \mathcal{A}_{j}^{i}\equiv \mathcal{Q}_{j}^{nm} y_{nm}  \mathcal{C }_{nm}^{i}.
\end{equation}
Note that we base our discussion here on the scalar problem for simplicity.  The vector problem can be formulated using the scheme presented in Sec.~\ref{SEC:vec}.

Here we first discuss timescales of the problem, and the general response to a dynamic signal will be discussed in the next subsection (Sec.~\ref{SEC:finiteW}).  To do this, we study the characteristic equation of these problems
\begin{align}
    Y(\omega) V(\omega) = \mathcal{A}(\omega) V(\omega), 
\end{align}
which correspond to the homogeneous solution to Eq.~\eqref{EQ:EOMW}.  Note that this is not yet a conventional eigenvalue problem, as $Y(\omega), \mathcal{A}(\omega)$ depend on $\omega$ in the general case, as we discussed above.  The general homogeneous solution involves a nonlinear equation of $\omega$.

Nevertheless, it is instructive to consider a few simple cases, where we have only one type of elements for the nodes and one type of elements for the edges.  This enables us to pull out the $\omega$ dependence, and turn the characteristic equation into a conventional eigenvalue problem.  
Let's take the mass-on-nodes and dashpot-on-edges problem as a simple example.  In this case, 
\begin{align}
    Y(\omega) &= i\omega M ,\\
    \mathcal{A}(\omega) &= \mathcal{A},
\end{align}
where $M$ is the constant matrix containing all node masses and $\mathcal{A}=\mathcal{Q}_{j}^{nm} \eta_{nm}  \mathcal{C }_{nm}^{i}$ is the constant matrix containing all edge dashpot coefficients (i.e., all edges are characterized by dashpots containing viscous fluids where force is proportional to velocity with coefficient $\eta_{nm}$).  The characteristic equation is then
\begin{align}
    i\omega V(\omega) = M^{-1} \mathcal{A} V(\omega),
\end{align}
determining that the homogeneous solution is a linear combination of exponential decay modes with decay rate determined by the eigenvalues of $M^{-1} \mathcal{A}$,
\begin{align}
    i \omega = \textrm{spec}(M^{-1}\mathcal{A}),
\end{align}
where $\textrm{spec}(M^{-1}\mathcal{A})$ denote the set of eigenvalues of $(M^{-1}\mathcal{A})$.  
This enables us to write the general homogeneous solution as
\begin{equation}
    V(t)=\sum_{\alpha} e^{-\lambda_{\alpha}t},
\end{equation}
where $\lambda_{\alpha}$ are eigenvalues of $M^{-1}\mathcal{A}$, labeled by $\alpha$. 

One can similarly write this homogeneous solutions for other cases.  For the dashpot-on-nodes and spring-on-edges problem, we have 
\begin{align}
    Y(\omega) &= H ,\\
    \mathcal{A}(\omega) &=\frac{1}{i\omega} \mathcal{K},
\end{align}
giving decay rates
\begin{align}
    i \omega = \textrm{spec}(H^{-1}\mathcal{K}).
\end{align}
The cases of dashpot-on-nodes and mass-on-edges, as well as spring-on-nodes and dashpot-on-edges, follow similar analysis, giving pure decay solutions. 

These  cases are characterized by exponential decay because the equation of motion involves only the first order time derivative.  We get oscillatory solutions when there are second order time derivatives, which shows up in two cases in this context.  The first one is the mass-on-nodes and spring-on-edges problem,
\begin{align}
    Y(\omega) &= i\omega M \\
    \mathcal{A}(\omega) &=\frac{1}{i\omega} \mathcal{K} ,
\end{align}
giving oscillatory frequency scales
\begin{align}\label{EQ:MSw}
    \omega = \sqrt{\textrm{spec}(M^{-1}\mathcal{K})},
\end{align}
which is a generalization of the familiar harmonic oscillator frequency $\sqrt{k/m}$. 

A similar case arises when we put masses on edges and springs on nodes,
\begin{align}
    Y(\omega) &= \frac{1}{i\omega} K \\
    \mathcal{A}(\omega) &= i\omega \mathcal{M},
\end{align}
giving oscillatory frequency scales
\begin{align}
    \omega = \sqrt{\textrm{spec}(\mathcal{M}^{-1}K)},
\end{align}
which shares a form that appears similar to Eq.~\eqref{EQ:MSw} but has important differences, as $\mathcal{M}$ describes edge masses and $K$ nodes spring constants. 

\begin{figure*}[t]
    \centering
    \includegraphics[width=0.95\textwidth]{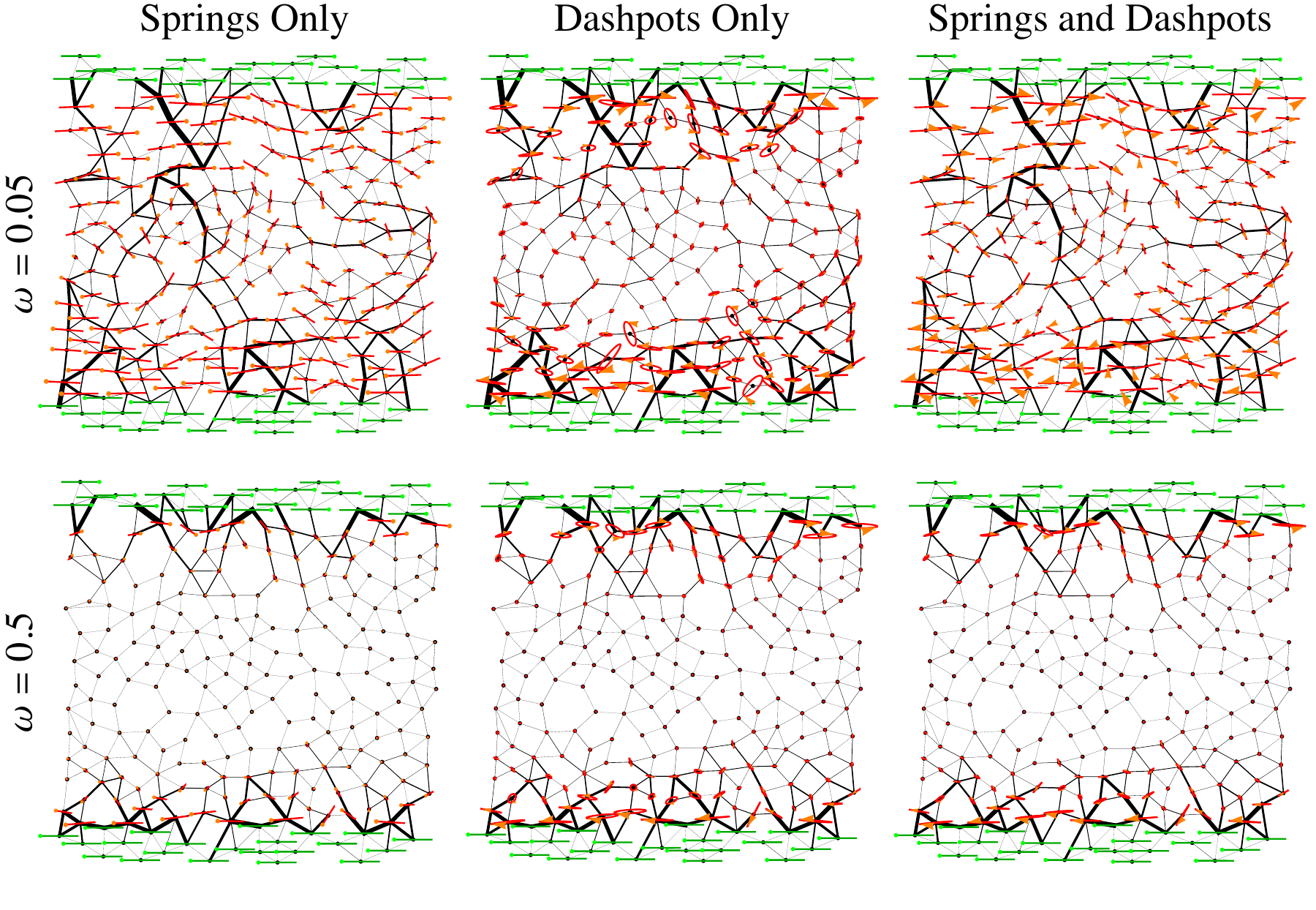}
    \caption{The dynamic response of spring only (left), dashpot only (middle), and spring with dashpot (right) networks to imposed dynamic displacements at frequencies $\omega= 0.05$ (top) and $\omega= 0.5$ (bottom).
    Applied displacements (dark green) are periodic horizontal oscillations to nodes at the top and bottom of the network.
    The osculations to the top of the network are exactly out of phase with those at the bottom of the network, shown by dots indicating the position of the node at an equal time (light green).
    The response displacements of nodes in the bulk (red), which in general are elliptical orbits (direction and equal time positions shown by orange arrow, orange dot when no eccentricity), penetrates further into the bulk of system at lower frequencies, more closely resembling the static displacement problem. 
    Edges (black lines) have thickness proportional to the magnitude of their complex tension response.}
    \label{fig:dynamicPanel}
\end{figure*}

Finally, we should also include the ``diagonal'' cases where both nodes and edges have only mass, only dashpot, and only springs.  These cases exhibit \emph{no timescales} and only trivial solution of $V=0$ for the homogeneous equation. A summary of these cases can be found in Tab.~\ref{mechanicalEOM}. Other types of problems on physical networks, such as electrical, heat and mass transport, can be calculated in similar ways. 

It is also worth noting that such eigenvalues analysis is enabled by assuming simple elements on the physical network.  The general case where $Y(\omega)$ and $\mathcal{A}(\omega)$ depend on $\omega$ in general ways can not be fully captured in such simple analysis.  One will need to solve the full equation.  

\subsection{Finite-frequency response of physical networks}\label{SEC:finiteW}
With these homogeneous solutions in place, we can now discuss the inhomogeneous solution to the general equation~\eqref{EQ:EOMW}.  
The solution can be formally written as
\begin{align}
    V(\Omega) = (Y(\Omega)I-\mathcal{A}(\Omega))^{-1} \tilde{S}(\Omega) ,
\end{align}
where $\tilde{S}(\Omega)=S(\Omega)+Q y(\Omega)w$ is the effective node sources including the edge driving, and $I$ is the identity matrix.  We use uppercase $\Omega$ to signify the signal frequency.  

It is helpful to think of this as a generalization of a driven damped harmonic oscillator problem, where the frequency domain response is
\begin{equation}\label{EQ:uw}
    u(\Omega) = \frac{f(\Omega)}{m\Omega^2-i\gamma\Omega-k},
\end{equation}
where $m,\gamma,k$ are the mass, drag (dashpot) coefficient, and spring constant of the oscillator.  The time domain response is then
\begin{equation}
    u(t) = \int d\Omega \, e^{-i \Omega t} \frac{f(\Omega)}{m\Omega^2-i\gamma\Omega-k}
\end{equation}
which is typically evaluated using contour integrals (adding a semicircle at $\Omega''\to -\infty$ for retarded Green's functions ($t>0$)), and thus 
governed by the poles of the frequency domain solution Eq.~\eqref{EQ:uw}.

Similarly, for physical networks, the calculation is similar, with the simple factor $\frac{1}{m\Omega^2-i\gamma\Omega-k}$ replaced by the network green's function $(Y(\Omega)I-\mathcal{A}(\Omega))^{-1}$, but the general formulation follows similarly.  The most nontrivial element here, in some sense, is the complex $\Omega$ dependence in both $Y(\Omega)$ and $\mathcal{A}(\Omega)$, causing nonlinear equations determining the poles.

We show some examples of this computation in Fig.~(\ref{fig:dynamicPanel}) where a mechanical network is driven at given frequencies at chosen boundary nodes.  

\section{Conclusion and outlook}
We have developed a comprehensive and unified framework for analyzing the linear response of physical networks---ranging from mechanical lattices to electrical circuits, thermal transport systems, and beyond---grounded in algebraic graph theory. This approach offers a mathematically rigorous yet versatile language that captures both static and dynamic behaviors, scalar and vectorial flows, and conservative as well as dissipative processes. By identifying flows and potentials on network edges and nodes, and systematically relating them via incidence and cut-set matrices, we derived a set of universal equations governing physical responses across domains.

Our formulation reveals deep analogies between seemingly disparate systems. For instance, mechanical force balance and thermal current conservation share algebraic structure through Kirchhoff's laws, while entropy production in irreversible systems mirrors elastic energy in static networks. Furthermore, our use of linear algebra—particularly the interplay of column spaces, null spaces, and dual graphs—unifies these analogies and suggests that many known results (such as self-stress states, floppy modes, and reciprocal diagrams) are manifestations of a more general structure.

Looking forward, this framework opens multiple avenues for exploration.  One immediate extension is  multiphysics~\cite{zimmerman2006multiphysics} on networks: The algebraic structure naturally accommodates multiple physical variables (e.g., stress, heat, mass, charge) and their couplings via generalized admittance or resistance matrices, providing a principled path to study cross-phenomena effects like thermoelectric or chemo-mechanical responses. 

While our work focuses on linear responses, the underlying network formalism is amenable to generalization. Incorporating nonlinear elements (e.g., threshold conductances, active feedback) or time-dependent control (e.g., space-time modulation) could extend this method to active matter, neuromorphic systems, and non-Hermitian physics.

Furthermore, the ability to condense nodes and describe macroscopic strain or stress from microscopic detail hints at a promising direction for renormalization-like procedures on networks~\cite{gabrielli2025network}. This could lead to systematic coarse-graining tools that preserve key physical responses across scales.

Beyond physics and engineering, our formulation resonates with recent developments in machine learning and artificial intelligence, particularly in graph-based models and network inference. The algebraic structure underpinning physical networks parallels the architecture of graph neural networks (GNNs)~\cite{1555942}, where information propagates along edges and is aggregated at nodes, opening a route for data-driven discovery of effective models and materials.  At the same time, these tools also provide a convenient basis for the study of physical neural networks, where intrinsic physical dynamics on networks are utilized for information processing~\cite{stern2021supervised,rocks2019limits,rocks2020revealing,stern2023learning,li2024training,li2025topological}.

\noindent{\it Acknowledgments.}---The authors thank Nick Kotov, Bulbul Chakraborty, Mark Newman, and Anthony Grbic for helpful discussions.  
This work was supported in part by 
the National Science Foundation Center for Complex Particle Systems~(Award \#2243104), National Science Foundation Award \#2026825, and 
the Office of Naval Research (MURI N00014-20-1-2479).

\appendix

\section{Graph Theory and Linear Algebra}\label{SEC:AppA}

Here we will review some basic concepts of linear algebra and graph theory as they apply to our present work. There appears to be much variation in the language and conventions previously used when discussing this topic. Here we make our own language and conventions clear. First we discuss the concept of cycles and cut sets \cite{strang2016introduction,Bollobás1998,swamy1981graphs,Lloyd1978GraphTW}. Then we discuss the fundamental theorem of linear algebra (FTLA)~\cite{fundamental_theorem_linalg}  and its consequences for physical networks. 
  
\subsection{Cycles and Cut Sets}

When studying transport on a network it is useful to focus on the edges, since this is where the transport takes place. Certain sets of edges are of fundamental importance. Here we will discuss cycles and cut sets.  A cut set is a set of edges such that if eliminated splits the graph into two connected components. A cycle is a set of edges that form a closed path. In what follows we will describe cycles and cut sets of a graph in linear algebra terms as vectors in $\mathbb{R}^{N_e}$ where $N_e$ is the number of edges in the graph.
    
Firstly we start with a undirected graph $G$ and choose some ordered list of its edges to represent it. We call this ordered list the \textbf{edge list}. For example, let $G$ be the graph with edge list  $(1,2),(1,3),(2,3),(2,4)$. This chosen ordering of the graph, for both edges and nodes within edges, is a convention and should have no effect on physics. Note that this ordering allows us to refer to edges by their index. For example edge number 3 is $(2,3)$. In terms of vectors we write $(2,3)$ as $\langle0,0,1\rangle$ and $(3,2)$ as $\langle0,0,-1\rangle$. In this manner we can represent any subset of edges and the orientation in which they are given. For example $(1,2),(3,1)$ would correspond to vector $\langle1,-1,0\rangle$.  To be more explicit the vector corresponding to a given edge subset $G'$ with is own edge list,  is obtained in the following way:  if the nth edge of $G$ is not on $G'$ the nth element of the vector is $0$, it is $1$ if it is on $G'$ and listed the same way and $-1$ if it is there but listed in reverse order.

In other presentations of this topic the graph is treated as a directed graph. Here we have chosen to use this object we call the edge list and continue to refer to $G$ as undirected for a number of reasons. First there is a conceptual reason, we want to make clear that the orientation induced on the graph by this ordered list will not correspond to any inherent property of our object of study, a physical network. The orientation is just a convention, similar to choosing a reference frame. There is also a practical computational aspect, which is that this ordered edge list allows us to build vectors and matrices and perform linear algebraic operations in a clear standardized way. This is has proven crucial in the process of writing code to numerically perform all the methods we will discuss.  
     
A \textbf{cycle} in a graph can be represented by a list of edges for example $(1,2),(2,3),(3,1)$, note how we organize the edges so that the second node of one edge is the first node of the next one in the list and the first and last nodes are the same. In this way we represent that we have taken a closed walk visiting the nodes in sequence $1,2,3,1$. The cycle, a subgraph $G'$ of our previously defined $G$ , described as an ordered list $(1,2),(2,3),(3,1)$ maps to the vector $\mathbf{w}=\langle 1,-1,1,0\rangle$.   We will use the word cycle to refer to the closed trail, the subgraph, and its vector representation interchangeably.    The linear subspace spanned by the cycles is called the \textbf{cycle space} and a basis for this space is called a \textbf{cycle basis}.

A \textbf{cut set} is a subset of the edges that when removed divides the graph into two disconnected parts. More formally a cut set is a set of edges such that there exists  a partition of a graph into two sets of nodes such that all edges in the set have one node in each element of the partition.  Now, consider $G$ our example graph $(1,2),(1,3),(2,3),(2,4)$. Note $G$ is connected, if we eliminate edges $(1,2)$ and $(2,3)$ we disconnect it into two components $G_a$ and $G_b$ whose nodes are $\{1,3\}$ and $\{2,4\}$. We have found a cut-set. Like in the case of cycles, cut sets have to be written with a consistent orientation of the edges. Cut sets should be written all pointing from component $G_a$ to $G_b$ (or all vice-versa). Our cut set is then $(1,2),(3,2)$, which maps to the vector $\langle1,0,-1,0\rangle$.  The linear subspace spanned by the ct sets is called the \textbf{cut space} and a basis for it is called a \textbf{cut basis}.
      
One can easily verify that our example cycle is orthogonal to our example cut set. This is in fact always the case. For a given graph all of its cut sets are orthogonal to all of its cycles. Since the cut space and the cycle space are \textbf{orthogonal complements}. This is a well established result and is related to the fundamental theorem of linear algebra.  We will not provide a proof. One way to intuitively see why this is the case is to consider a cycle and a cut set which are not disjoint. The cycle must intersect the cut set an even number of times. Say the cut set separates the graph into components $G_a$ and $G_b$. The cycle must exit $G_a$ the same number of times it enters $G_a$. In each exit the cycle is oriented along the cut set but on each entry its oriented opposite to the cut set. These naturally makes the two vectors orthogonal.

Now, one can relate a number of matrices with a graph $G$. It is common to use the adjacency matrix as a representation of a graph. One inconvenience of this approach is that the adjacency matrix goes from the nodes to the nodes. Since we are interested in studying what goes on in edges we will need different matrices.  Let us introduce the incidence matrix $\mathcal{C}$  which in our convention has as many rows as there are edges and as many columns as there are nodes. Here we will also make use of an ordered edge list for the graph.  
Let $\text{EdgeList}(n)$ refer to the ordered pair that is the nth element of the edge list. 
We then write:
\begin{align}
    \mathcal{C}_{n,i} \,= \begin{dcases}
        1,&   \text{if}\, (i,j) = \text{EdgeList}(n),\exists j\\
        -1,&   \text{if}\, (j,i) = \text{EdgeList}(n),\exists j \\
        0,& \text{otherwise}. \\
    \end{dcases} 
\end{align}
The column space of this matrix is the cut set of the graph. 

Another relevant matrix is the cycle matrix $\mathcal{B}$, (also circuit matrix) which for us is any matrix such that its rows form a complete cycle basis.  
Traditionally this matrix is more narrowly defined by requiring that all rows correspond to cycles as we have described them before. 
This results in all entries being 0,1 or -1.
Since the cycle space and cut space are orthogonal complements:
\begin{equation}
    0=\mathcal{B}\mathcal{C}\mathbf{x}  \, ,\forall \mathbf{x}. 
\end{equation}
Further, since the two spaces are complements, any vector $\mathbf{w}$ in the space of edges can be represented as a sum of a cut set component and a cycle component:
\begin{equation}
    \mathbf{w}=  \mathcal{C}\mathbf{x}  +  \mathcal{B}^T \mathbf{\tilde{x}}.
\end{equation}

\subsection{Fundamental Theorem of Linear Algebra}
   
Here we will discuss the fundamental theorem of linear algebra and explain what it tells us about cycles and cut sets of a graph. 
The fundamental theorem of linear algebra states \cite{fundamental_theorem_linalg}: Given a real valued $m\times n$ matrix $A$, 

\begin{enumerate}
    \item $\text{dim } \text{col}(A) = \text{dim } \text{col}(A^T)$  and, $\text{dim} \, \text{col}(A) + \text{dim}  \,\text{null}(A^T) = m$,
    \item $\text{null}(A^T)$ and $\text{col}(A)$ are orthogonal complements.
\end{enumerate}    

Let us apply the FTLA to the incidence matrix. 
We know that $\text{col}(\mathcal{C})$ is the cut space. 
The FTLA then tells us that $\text{null}(\mathcal{C}^T)$ is the orthogonal complement of the cut space: the cycle space,
\begin{equation}
    \text{null}(\mathcal{C}^T)\equiv \, \text{col}(\mathcal{B}^T).
\end{equation}
    
In other words:
\begin{equation}
    0=\mathcal{C}^T\mathcal{B}^T\mathbf{\tilde{x}} \,\,,\forall  \mathbf{\tilde{x}}.
\end{equation}
    
Part one of the FTLA theorem is also known as the rank nullity theorem from which one can easily show that for any real valued   $m\times n$  matrix , say $\mathcal{C}$:    
\begin{equation}
    \text{dim null}(\mathcal{C}^T)-  \text{dim null}(\mathcal{C}) =  m -   n. \label{eq:fundamental_counting}
\end{equation}
One might call this a fundamental counting rule. If $\mathcal{C}$ is the incidence matrix of a given graph, then: 
\begin{itemize}
    \item $\text{dim null}(\mathcal{C}^T)= N_{\text{cycles}}$ : The dimension of the cycle space.
    \item $\text{dim null}(\mathcal{C}) = N_{\text{c.c.}} $: The number of connected components.
    \item $m= N_{\text{edges}}$.
    \item $n= N_{\text{nodes}}$.
\end{itemize}
    
The counting rule then gives us the following relationship:
\begin{equation}
    N_{\text{cycles}}-N_{\text{c.c.}}=N_{\text{edges}}-N_{\text{nodes}} , \label{maxwellc_graphs}
\end{equation}
note that $N_{\text{c.c.}}$ (the number of connected components) is $1$ for a completely connected graph. 
We will later see that this is a case of what is often referred to as  \textbf{Maxwell's counting}~\cite{CALLADINE1978161}.\\

This counting relationship could also be written in terms of the cut space by making use of the fact that because it is the orthogonal complement to the cycle space its dimension $N_{\text{cut}}$ satisfies
\begin{equation}
    N_{\text{edges}}=N_{\text{cycles}}+N_{\text{cut}}.
\end{equation}   
    
\subsection{General Least Squares Minimization Problem}

Consider the following problem.  Given a real valued $m\times n$  matrix  $\mathcal{C}$  and an $m\times 1$ vector $\mathbf{w}$ find $ \mathbf{w}-\mathcal{C}\mathbf{x}$ such that  $ |\mathbf{w}-\mathcal{C}\mathbf{x}|$ is minimized . We will see that we can cast a variety of physical network problems into this form and solve them by the same generalized approach. 

This problem can be solved by using the Moore Penrose pseudo inverse $\mathcal{C}^{+}$,
\begin{equation}
    \mathbf{x}_{min}=\mathcal{C}^{+}\mathbf{w}.
\end{equation}
    
The matrix $\mathcal{C}^{+}$ is obtained from the singular value decomposition.  
The solution to our problem is then: 
\begin{equation}
    \mathbf{w}-\mathcal{C}\mathbf{x}_{min}=\left(1-\mathcal{C}\mathcal{C}^{+}\right)\mathbf{w}.
\end{equation}
       
Let us explore the problem from a different perspective. 
It is clear that $ \mathbf{w}-\mathcal{C}\mathbf{x}=0$ only has solution if $\mathbf{w}$ is in the column space of $\mathcal{C}$. 
From a geometric point of view the closest $\mathcal{C}\mathbf{x}$ can get to $ \mathbf{w}$ is the projection of $ \mathbf{w}$ along the column space of $\mathcal{C}$. 
The minimum difference is then given by the projection into the orthogonal complement of $\mathcal{C}$ (See Fig.~(\ref{fig:edge_swelling_projection})):
\begin{equation}
    \mathbf{w}-\mathcal{C}\mathbf{x}_{min}=\hat{\mathcal{B}}^T\hat{\mathcal{B}} \mathbf{w},
\end{equation}
where $\hat{\mathcal{B}}$ is a matrix such that its rows form an orthonormal basis for $null(\mathcal{C}^T)$.
We can alternatively define a matrix $\hat{\mathcal{C}}$ such that its columns are an orthonormal basis for $col(\mathcal{C})$.
One suitable choice for this basis are the left singular vectors of $\mathcal{C}$. 
We can then write:
\begin{equation}
    \mathbf{w}-\mathcal{C}\mathbf{x}_{min}=\left(1-\hat{\mathcal{C}}\hat{\mathcal{C}}^T\right)\mathbf{w}.
\end{equation}

\section{Other Electrical-Mechanical Analogies}\label{SEC:AppB}
As mentioned in the main text, there are a number of ways one can map electrical circuits to mechanical ones and vice versa. 
In addition to Firestone's analogy~\cite{Firestone} discussed in the main text, here we  will discuss the more conventional impedance analogy,sometimes attributed to Maxwell~\cite{Busch-Vishniac_1999}, and our own original analogy. 

For the discussion of these analogies it is convenient to talk in terms of impedance $z$ which is the reciprocal of the admittance $z=1/y$. ``Mechanical impedance" is defined differently for each analogy. We will present a number of equations of motion all analogous to Eq.~(\ref{EQ:EOMW}).

\subsection{Impedance Analogy: force as potential, velocity as current}
The impedance analogy is the oldest and more conventional analogy~\cite{Firestone}.  
It maps electrical impedance to mechanical impedance defined as the frequency dependent complex quantity that relates velocity to force as:
\begin{equation}
    f=z\dot{u},
\end{equation}
where $\dot{u}$ is the node velocity and $f$ the force on a node. 
Similarly, edge impedance is the coefficient that relates the rate of change of the elongation and edge tension:
\begin{equation}
    t=z\dot{e},\label{eq:imp1}
\end{equation}
while electrical impedance is the coefficient that relates voltage and current:
\begin{equation}
    V=zI.\label{eq:imp2}
\end{equation}

\begin{figure}[h]
    \centering
    \includegraphics[width=0.95\linewidth]{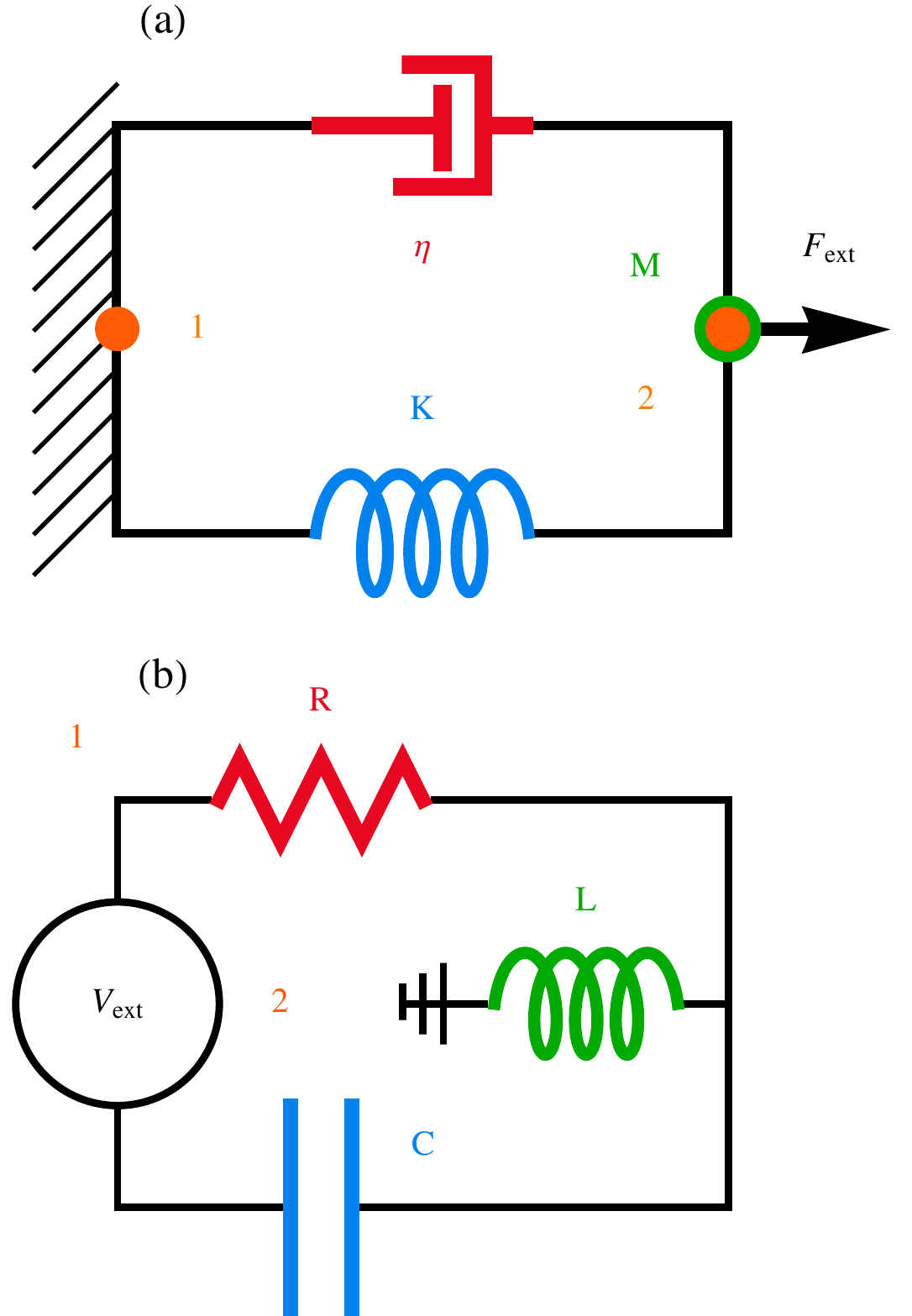}
    \caption{Analogous components (same color/position) between mechanical (a) and electrical (b) networks as described by the Impedance Analogy. 
    Note that the topology of these two networks is different.
    In the mechanical analogy, the dashpot and spring are in paralell, but their analogous components in the electrical network (resistor and capacitor) are in series.
    }
    \label{fig:impedanceAnalaogy}
\end{figure}

The Impedance Analogy is based on the analogy between Eqs.~(\ref{eq:imp1},\ref{eq:imp2}). 
Voltage maps to tension, electric potential to force, the rate of elongation maps to edge currents and node velocity maps to current sources. 
This mapping is opposite to the electrical-mechanical analogy we used in previous sections for the static case.
The impedance analogy has a serious drawback which is that the mechanical circuit analogous to a given mechanical circuit does not have the same topology (as shown in Fig.~\ref{fig:impedanceAnalaogy}). 
Elements that are in parallel in one are in series in the other and vice versa. 
More generally under the impedance analogy cycles in the electrical circuit map to cut sets in the mechanical and vice versa. 
This means that the graphs underlying the two circuits are the dual of each other. 

\begin{table*}[t]\label{impAnalogyTab}
\begin{tabular}
{|c|c|c|c|}
  \hline
  $\quad$Mechanical Elements\rule{0pt}{3ex}\rule[-2ex]{0pt}{0pt} $\quad$ & $\quad$Mechanical Impedance$\quad$ & $\quad$Electrical Elements$\quad$ & $\quad$Electrical Impedance$\quad$ \\
  \hline
  Mass/Inertia\rule{0pt}{3ex}\rule[-2ex]{0pt}{0pt} & $i \omega M$ & inductor & $i \omega L$ \\
  \hline
  Dashpot\rule{0pt}{3ex}\rule[-2ex]{0pt}{0pt} & $\eta$ & resistor & $R$ \\
  \hline
  Spring\rule{0pt}{3ex}\rule[-2ex]{0pt}{0pt} & $K/i\omega$ & capacitor & $i/i\omega C$ \\
  \hline
\end{tabular}
\caption{Impedances of mechanical and electrical elements in the \textbf{Impedance Analogy}.}
\end{table*}

We can see this duality at play when we consider a  1D spring network initially unstressed and then driven at a frequency $\omega$.  
The elongation rates will be compatible just like the elongations themselves meaning that they add to zero along cycles:
\begin{equation}
    0=\sum_{ij\in \text{cycle}} \dot{e}_{ij}.
\end{equation}
For the electric analogous electric circuit we will have the current conservation law:
\begin{equation}
    0=\sum_{ij\in \text{cut set}} I_{ij}.
\end{equation}
In order for these two conditions map to each other under the impedance analogy cycles must map to cut sets and vice versa.

Our general equation Eq.~(\ref{EQ:EOMW}) for mechanical systems, with impedance defined as  $z=t/\dot{e}$ would read
\begin{equation}
    \Lambda^{-1}\dot{\mathbf{u}}= -QZC\dot{\mathbf{u}}+QZ\dot{\mathbf{e}}_0+ \mathbf{f}_0
\end{equation}
Here $Z$ and $\Lambda$  are the diagonal matrices of mechanical impedances of edges and nodes respectively. In the context of Eq.~(\ref{EQ:EOMW}) $Y=\Lambda^{-1}, \mathcal{A}=QZC$.  Note that although in the impedance analogy $\dot{\mathbf{u}}$ is formally analogous to current it appears in the equation in the same way as electrical potential. From an algebraic perspective one would more naturally arrive at a different analogy where velocity is potential. This is precisely Mobility analogy we will discuss next.

One could alternatively write the equation of motion as:
\begin{equation}
    \Lambda\mathbf{f}= -\tilde{Q}Z^{-1}\tilde{C}\mathbf{f}+\tilde{Q}Z^{-1}\mathbf{t}_0+ \mathbf{u}_0,
\end{equation}
where $\tilde{C}$ is the incidence matrix (or compatibility matrix) for the dual graph. 
Note that this would be the more natural way of writing this equation according to the impedance analogy where force plays the role of potential.

\subsection{Firestone's Analogy: velocity as potential, force as current}

In 1933, Firestone, noticing the problem of the impedance analogy not being topology preserving on a network, proposed a different analogy \cite{Firestone}. 
This analogy is sometimes called the admittance analogy since it maps electrical elements to mechanical ones in such a way that the electrical impedance matches the mechanical admittance (reciprocal of impedance). 
For clarity, we refer to this analogy as the ``Firestone" analogy or ``Firestone's" analogy, which we discussed significantly in the main text.
Firestone's analogy preserves topology and is more in line with the theory we have presented in this work. 
In his original paper, Firestone went as far as to strongly propose that mechanical impedance be redefined to $\dot{u}/f$ .  
For this reason Firestone called the mechanical admittance the bar impedance $\bar{z}$  a notation we will make use of here. 
Mechanical admittance is also called mobility and so the analogy is also sometimes called the mobility analogy. 

Mobility $\bar{z}$ is the coefficient that relates forces and displacements as:
\begin{equation}
    \dot{u}=\bar{z} f,
\end{equation}
where $\dot{u}$ is the node velocity and $f$ the force on a node. Similarly, edge mobility is the coefficient that relates the rate of change of the elongation and the tension:
\begin{equation}
    \dot{e}=\bar{z}\,t.
\end{equation}
When we map the mobility to electrical impedance we find that tension maps to current, force on nodes to current sources, elongation rate to voltage and velocity to electric potential. 
Now elongation rates that sum to zero along cycles map to voltages that sum to zero along cycles.

In this analogy, charge maps to momentum, therefore charge flow (current) maps to momentum flow (force). 
The analogies between the different elements can be explained conceptually through this idea. 
For example, in the same way a capacitor stores charge a mass or a inertia stores momentum. 

In terms of bar impedance we can write the equation of motion as:
\begin{equation}
    \bar{\Lambda}\dot{\mathbf{u}}= -Q\bar{Z}^{-1}C\dot{\mathbf{u}}+Q\bar{Z}^{-1}\dot{\mathbf{e}}_0+ \mathbf{f}_0.
\end{equation}
Note how this equation is essentially identical to the electrical case.  In the context of Eq.~(\ref{EQ:EOMW}) $Y=\bar{\Lambda}, \mathcal{A}=Q\bar{Z}^{-1}C$.

\begin{table*}[t]\label{firestoneAnalogyTab}
\begin{tabular}
{|c|c|c|c|}
  \hline
  $\quad$Mechanical Elements\rule{0pt}{3ex}\rule[-2ex]{0pt}{0pt}$\quad$ & $\quad$Mechanical Mobility$\quad$ & $\quad$Electrical Elements$\quad$ & $\quad$Electrical Impedance$\quad$ \\
  \hline
  Mass/Inertia\rule{0pt}{3ex}\rule[-2ex]{0pt}{0pt} & $1/i\omega M$ & Capacitor & $1/i \omega C$ \\
  \hline
  Dashpot\rule{0pt}{3ex}\rule[-2ex]{0pt}{0pt} & $1/\eta$ & Resistor & $R$ \\
  \hline
  Spring\rule{0pt}{3ex}\rule[-2ex]{0pt}{0pt} & $i \omega/K$ & Inductor & $i \omega L$ \\
  \hline
\end{tabular}
\caption{Mobilities and impedances of mechanical and electrical elements in the \textbf{Firestone Analogy}.}
\end{table*}

\subsection{``Static Consistent'' Analogy: displacement as potential, force as current}

\begin{figure}[h]
    \centering
    \includegraphics[width=0.95\linewidth]{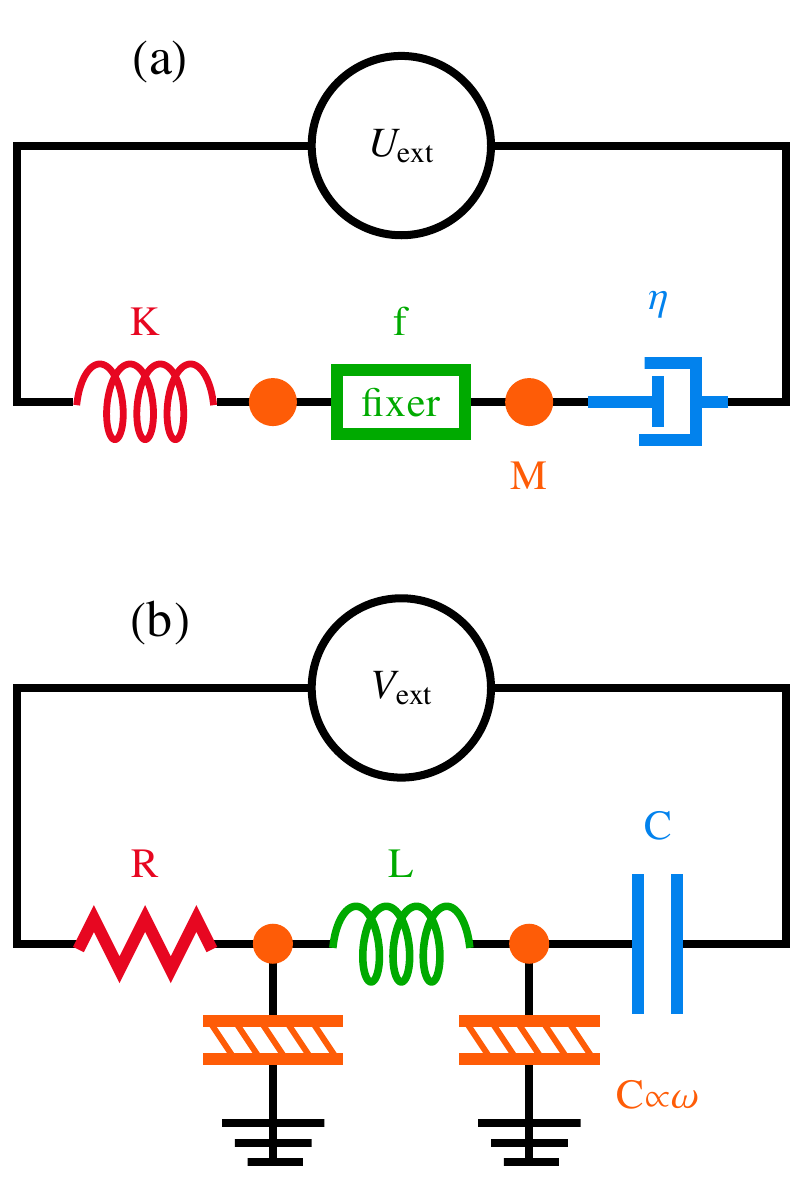}
    \caption{Analogous components (same color/position) between mechanical (a) and electrical (b) networks as described by the Static Consistent Analogy. 
    Note that the node-capacitors (analogous to mass) must have a capacitance that is proportional to frequency.
    }
    \label{fig:consistant}
\end{figure}

In this work we previously presented a mechanical-electrical analogy between static equilibrium and a constant current steady state. 
We identified the node displacement with the electric potential. This does not agree with either the impedance or mobility analogy, it is, in fact a third analogy. 
The usefulness of the ``displacement as potential" analogy lies in the fact that it holds for both static and dynamic problems where as the other analogies do not apply in the static case. 
For this reason, we call this newly introduced analogy the ``Static Consistent'' Analogy, as it is consistent between the static and dynamic cases.
Fig.~(\ref{fig:consistant}) shows the analogous elements of this analogy, note that springs and resistors are analogous here, even with time-dependence.

We define our stand in for the impedance as $\tilde{z}$ which satisfies:
\begin{equation}
   u=\tilde{z} f 
\end{equation}
for nodes, and
\begin{equation}
    e=\tilde{z}\,t
\end{equation}
for edges. 
The quantity $\tilde{z}$ is a generalized ``compliance'' (reciprocal of stiffness) in the same way impedance is a generalized resistance.  

\begin{table*}[t]\label{consistentAnalogyTab}
\begin{tabular}
{|c|c|c|c|}
  \hline
  $\quad$Mechanical Elements\rule{0pt}{3ex}\rule[-2ex]{0pt}{0pt}$\quad$ & $\quad$Mechanical Compliance ($\tilde{z}$) $\quad$& $\quad$Electrical Elements$\quad$ & $\quad$Electrical Impedance$\quad$ \\
  \hline
  Mass/Inertia\rule{0pt}{3ex}\rule[-2ex]{0pt}{0pt} & $1/\omega^2M$ & Capacitor* & $1/i \omega C, C \propto \omega$ \\
  \hline
  Dashpot\rule{0pt}{3ex}\rule[-2ex]{0pt}{0pt} & $1/i \omega \eta$ & Capacitor & $1/i\omega C$ \\
  \hline
  Spring\rule{0pt}{3ex}\rule[-2ex]{0pt}{0pt} & $1/K$ & Resistor & $R$ \\
  \hline
  ``Fixer''\rule{0pt}{3ex}\rule[-2ex]{0pt}{0pt} & $i \omega f$ & Inductor & $i \omega L$ \\
  \hline
\end{tabular}
\caption{Mobilities and impedances of mechanical and electrical elements in the \textbf{Static Consistent Analogy}.}
\end{table*}

An edge mass/inertia maps to a capacitor such that the capacitance depends linearly on the frequency. 
Such an element is not so unrealistic since the electrical permittivity is frequency dependent, so all real capacitors that make use of dielectrics have frequency dependent capacitances.  
A capacitor with capacitance $c=c_0 +c_\omega \omega$ can be represented as a dashpot and an edge mass/inertia connected in parallel. 

The ``fixer" is an active element that keeps increasing or decreasing the tension as long as it is not at its rest length. The corresponding equation for this element is:
\begin{equation}
    \dot{t}= -\eta \,e.
\end{equation}
It is worth noting that this is a mechanical element with memory. The value of tension depends on the past states of the edge:
\begin{equation}
    t_T=t_0+\int_0^T -\eta \, e \,dT. 
\end{equation}

One peculiar aspect of this analogy is that it maps conservative electrical elements to non conservative mechanical elements. 
For example, capacitors (conservative) are mapped to dashpots (dissipative). 
This analogy, when applied to dynamics, gives us a way to map a family of active mechanical systems to passive electrical ones. 

The equation of motion for this analogy now written as:
\begin{equation}
    \tilde{\Lambda}\mathbf{u}= -Q\tilde{Z}^{-1}C\mathbf{u}+Q\tilde{Z}^{-1}\mathbf{e}_0+ \mathbf{f}_0.
\end{equation}
In the context of Eq.~(\ref{EQ:EOMW}) $Y= \tilde{\Lambda}, \mathcal{A}=Q\tilde{Z}^{-1}C$.

\bibliographystyle{apsrev4-1}
\bibliography{references}

\end{document}